\documentclass[a4paper,reqno]{amsart}

\usepackage{amsmath}
\usepackage{amssymb}
\usepackage{graphicx}
\usepackage{latexsym}
\usepackage{amsfonts}
\usepackage[marginratio=1:1,top=3.5cm]{geometry}

\begin{document}

\title[Singularity confinement and full-deautonomisation]{Singularity confinement and full-deautonomisation: \\ a discrete integrability criterion}

\author{
B. Grammaticos$^{1}$, A. Ramani$^{2}$, R. Willox$^{3}$, T. Mase$^{3}$ and J. Satsuma$^{4}$}

\address{$^{1}$IMNC, Universit\'e Paris VII \& XI, CNRS, UMR 8165, B\^at. 440, 91406 Orsay, France\\
$^{2}$Centre de Physique Th\'eorique, Ecole Polytechnique, CNRS, 91128 Palaiseau, France
$^{3}$Graduate School of Mathematical Sciences, the University of Tokyo, 3-8-1 Komaba, Meguro-ku, 153-8914 Tokyo, Japan
$^{4}$Department of Mathematical Engineering, Musashino University, 3-3-3 Ariake, Koto-ku, 135-8181 Tokyo, Japan
}


\begin{abstract}
We present a new approach to singularity confinement which makes it an efficient and reliable discrete integrability detector. Our method is based on the full-deautonomisation procedure, which consists in analysing non-autonomous extensions of a given discrete system obtained by adding terms that are initially absent, but whose presence does not alter the singularity pattern. A justification for this approach is given through an algebro-geometric analysis. We also introduce the notions of early and late confinement. While the former is a confinement that may exist already for the autonomous system, the latter corresponds to a singularity pattern longer than that of the autonomous case. Late confinement will be shown to play an important role in the singularity analysis of systems with non-trivial gauge freedom, for which the existence of an undetected gauge in conjunction with a sketchy analysis, might lead to erroneous conclusions as to their integrability. An algebro-geometric analysis of the role of late confinement in this context is also offered. This novel type of singularity confinement analysis will be shown to allow for the exact calculation of the algebraic entropy of a given mapping. 
\end{abstract}

\maketitle

{\Small PACS numbers: 02.30.Ik, 05.45.Yv}

\section{Introduction}
Singularity confinement \cite{sincon} is a discrete analogue of the Painlev\'e property \cite{pprop} of ordinary differential equations, which infers the integrability of a given equation from the local structure of its singularities. The crucial requirement there is that singularities, the position of which depends upon the initial conditions, do not introduce multivaluedness (which in general makes it impossible to represent the solution of the differential equation as a function). Analogously, the singularity confinement approach is based on the local study of the singularities that appear in a discrete system. Here as well we are interested in singularities with positions that depend on the initial conditions of the system and singularity confinement requires those singularities to disappear after a few iteration steps, lest they lead to indeterminacies that make the construction of the solution of the system impossible. The relevance of singularity confinement  as an integrability detector is strengthened by the fact that all discrete systems integrable through spectral methods, studied to date, have been shown to possess confined singularities. On the other hand, linearisable discrete systems in general do not satisfy the singularity confinement criterion \cite{mimura}, in close parallel to what happens in the continuous case, where linearisable differential systems in general do not possess the Painlev\'e property either \cite{trembl}. 

If this parallel between singularity confinement and the Painlev\'e property had been perfect, we would of course have been in possession of an efficient and convenient discrete integrability detector. However, the discovery of non-integrable systems with confined singularities called the usefulness of singularity confinement as an integrability criterion into question. The best-known example of such a mapping is the one proposed by Hietarinta and Viallet in \cite{hiv}, which we refer to as the H-V mapping:
\begin{equation}
x_{n+1}+x_{n-1}=x_n+{1\over x_n^2}.\label{eqi}
\end{equation}
The pattern of its singularities is $\{x_{n+1}=0, x_{n+2}=\infty, x_{n+3}=\infty, x_{n+4}=0\}$ and, since $x_{n+5}=x_{n}$, its singularity is confined. However the authors of \cite{hiv} have shown numerically that this mapping exhibits large scale chaos and thus cannot be expected to be integrable. This of course raises the question what singularity confinement might mean in this case? Clearly, the confinement property is related to some subtle cancellations occurring when one iterates a rational mapping. These cancellations will in fact reduce the growth of the degree of the successive iterates. (We should point out here that, as shown by Bellon and Viallet \cite{bellon}, while the degree itself is not invariant under coordinate changes its growth is invariant and thus characteristic for the mapping). In fact, when a mapping is integrable, these confinement-related cancellations slow down the degree growth to such an extent that, asymptotically,   it becomes polynomial \cite{ohta}. However, in the case of the H-V mapping, whereas some cancellations do take place, these do not suffice to curb the asymptotic degree growth which remains exponential. Such rapid growth is the signature of non-integrability. Hietarinta and Viallet \cite{hiv} therefore introduced a quantitative measure of the degree growth of a rational mapping: its algebraic entropy. If $d_n$ represents the homogeneous degree of the numerator or denominator of $x_n$, the algebraic entropy of the mapping is given by the limit ${\mathcal E}=\lim_{n\to\infty}{1\over n}\log d_n$. For an integrable mapping the algebraic entropy must vanish. On the other hand, a non-zero value for ${\mathcal E}$ implies exponential growth and is therefore an indication of non-integrability. In the case of mapping \eqref{eqi} the algebraic entropy can be computed exactly \cite{take} and is found to be ${\mathcal E}=\log({3+\sqrt 5\over 2})$. 

Curiously, this mapping remained essentially a singleton as far as counterexamples to the confinement criterion were concerned (but see \cite{viallet}), despite the fact that the authors of \cite{hiv} presented general arguments for the existence of whole families of confining, non-integrable, mappings. The status of singularity confinement became even more complicated with the issue of late confinement \cite{late}. While standard practice in the implementation of singularity confinement, for mappings with parametric freedom, had been to enforce confinement at the very first possibility, it was not at all clear at the time why one should abide by this rule and why, for example, one could not postpone confinement until a later occasion. It turns out \cite{late},\cite{prsa} that when one opts for a late confinement, the resulting system will be non-integrable despite its singularities being confined (and despite the fact that when confinement is implemented normally, the resulting system might be integrable).

These problems led to a certain distrust of singularity confinement  as a method for detecting or deriving discrete integrable systems. Still, there has always existed a domain -- for which we coined the term deautonomisation \cite{capel} -- where this criterion continued to thrive and in fact furnished a slew of novel results. What we mean by {\sl deautonomisation}, is to consider the free parameters of a mapping (which a priori take constant values) to be functions of the independent variable, the precise form of which has to be obtained through the use of a certain discrete integrability criterion. The rationale behind this approach lies in the relation the growth properties of the solution of a mapping bear to its integrability. In the deautonomisation procedure one starts from an integrable autonomous system, obeying the low-growth requirement, and one seeks to extend it to a non-autonomous form while keeping the same growth. In most practical applications however, the integrability criterion one uses is in fact singularity confinement. The reason being that, compared to techniques that rely on the calculation of the algebraic entropy, the confinement criterion has the immense advantage that one can examine each singularity separately, establishing the constraints on the parameters one at a time and not all at once  in a hopelessly entangled way.

It is precisely this very same deautonomisation approach that will be shown to reinstate singularity confinement  as a reliable discrete integrability criterion. In \cite{redeem} we introduced the so-called {\sl full-deautonomisation} approach, and we claimed that this is the proper way to perform the singularity analysis of a given mapping. 
If the system is integrable, the characteristic equations for the constraints that one obtains for the parameters, will only have roots with modulus 1, whereas the presence of a root with modulus greater than 1 implies non-integrability. In  what follows we shall first illustrate the power of this approach through several examples, after which shall give a detailed discussion of the problems that arise due to gauge freedom in the mapping and of the solution the concept of late confinement offers to this conundrum.

\section{The full-deautonomisation procedure}\label{sec2}
The deautonomisation procedure consists in assuming that the parameters that appear in a mapping are functions of the independent variable and in using some integrability criterion, like singularity confinement, to fix their precise form. Standard practice when applying this procedure is to require that the (confined) singularity pattern of the autonomous mapping and that of its non-autonomous extension be identical. (An analogous requirement can be formulated whenever the algebraic entropy criterion is used: one then requires that the degree growths are the same for the autonomous and non-autonomous mappings). The full-deautonomisation procedure is an extension of the standard one, where one introduces terms (with non-autonomous coefficients) that are absent in the initial mapping but which, when present, do not modify the singularity pattern. To put it in a na\"\i ve but evocative way: one not only replaces 1s by functions, but also selected 0s.

We shall illustrate this procedure and its implications for the integrability of a given mapping on two examples. The first one is a mapping of the form
\begin{equation}
x_{n+1}+x_{n-1}={1\over x_n^2}.\label{eqii}
\end{equation}
Its singularity pattern is $\{0, \infty^2, 0\}$,  where by $\infty^2$ we mean that if we introduce a small quantity $\epsilon$ and assume that $x_n$ is finite and $x_{n+1}=\epsilon$, then $x_{n+2}$ will be of order  $1/\epsilon^2$. Deautonomising \eqref{eqii} then consists in replacing the numerator of the right-hand side by a function $a_n$ and to require the mapping to have confined singularities with exactly the same pattern as the autonomous one. This yields the constraint $a_{n+1}=a_{n-1}$, which gives an integrable, but trivial, non-autonomous extension of \eqref{eqii}. In fact, by introducing the appropriate gauge $x_n\to\gamma_n x_n$, with $\gamma_n^3=a_n^2/a_{n-1}$ and $\gamma_{n+1}=\gamma_{n-1}$, we can put $a_n=1$ for all $n$.

In order to proceed to the full-deautonomisation of \eqref{eqii} we must add terms that do not modify the initial singularity pattern. It is straightforward to convince oneself that the only possible such extension is by adding a term inversely proportional to $x$, which leads to
\begin{equation}
x_{n+1}+x_{n-1}={b_n\over x_n}+{a_n\over x_n^2}.\label{eqiii}
\end{equation}
Requiring the initial singularity pattern to be confined -- i.e. if we start from $x_{n+1}=\epsilon$, that $x_{n+4}$ takes a finite value which depends on $x_n$ -- we find the constraints $a_{n+1}=a_{n-1}$ and $b_{n+1}-2b_n+b_{n-1}=0$. The integration of the latter leads to the characteristic equation $(\lambda-1)^2=0$ and finally to $b_n=\alpha n+\beta$. This expression, combined with $a_n=1$, turns \eqref{eqiii} into a well-known form of the discrete Painlev\'e I equation \cite{capel}.

The second example is a non-integrable mapping with confined singularities of a form similar to \eqref{eqii},
\begin{equation}
x_{n+1}+x_{n-1}={1\over x_n^4},\label{eqiv}
\end{equation}
but with singularity pattern $\{0, \infty^4, 0\}$. Its standard deautonomisation follows exactly that of \eqref{eqii} and leads to the same constraint, and the same uninteresting result. Its full-deautonomisation is however of particular interest. Including in \eqref{eqiv} all the terms that leave the singularity pattern unchanged, leads to the mapping
\begin{equation}
x_{n+1}+x_{n-1}={b_n\over x_n}+{c_n\over x_n^2}+{d_n\over x_n^3}+{a_n\over x_n^4}.\label{eqv}
\end{equation}
The confinement constraints obtained by requiring that $x_{n+4}$ be finite, just as in the integrable case above, are: $a_{n+1}=a_{n-1}$, $d_{n+1}=-d_{n-1}$, $c_{n+1}=c_{n-1}$ and $b_{n+1}-4b_n+b_{n-1}=0$. The first three equations can be integrated in terms of periodic functions,  but they can also simply be satisfied by the choice $a_n=1$, $d_n=0$, $c_n=1$ or even $c_n=0$, without influencing the reasoning on the fourth constraint. Solving the latter, we find the characteristic equation $\lambda^2-4\lambda+1=0$ for which, contrary to the integrable case above, the roots are not roots of unity but are equal to $2\pm\sqrt 3$. As we posited in \cite{redeem}, the existence of a root in the characteristic equation for the coefficients, with modulus greater than 1, is an indication of non-integrability. Our argument is further strengthened by the observation that the logarithm of the largest root is precisely the value of the algebraic entropy that can be computed using the standard numerical procedure.

Thus our claim is that the proper way to perform the singularity analysis using the singularity confinement criterion is by performing a full-deautonomisation of the mapping. While our arguments may appear heuristic at this stage it is possible to make them more rigorous by a detailed  algebro-geometric analysis of these mappings.

\section{An algebro-geometric justification of full-deautonomisation}
In this section we analyse the mappings \eqref{eqiii} and \eqref{eqv} from the point of view of algebraic geometry. In particular, we shall derive conditions on the parameters, by blowing up the mappings at their indeterminate points, such that, ultimately, each mapping induces an automorphism on an appropriately structured rational surface, obtained by blowing-up $\mathbb{P}^1 \times \mathbb{P}^1$. This requirement may well be stronger than the singularity confinement criterion itself (we plan to clarify this point in a future publication) but it will be shown that the conditions on the parameters we obtain from this analysis do coincide with those obtained in the previous section. Moreover, the role these conditions play in determining the algebraic entropy of a mapping will also become clear.

\subsection{The case of an integrable mapping}
Let us start with \eqref{eqiii}, which we shall interpret as the following birational map on $\mathbb{P}^1 \times \mathbb{P}^1$:
\begin{equation}
\varphi_n \colon\quad \mathbb{P}^1 \times \mathbb{P}^1 \dashrightarrow \mathbb{P}^1 \times \mathbb{P}^1,
	\qquad (x_n, y_n) \mapsto (x_{n+1}, y_{n+1}) = \left( y_n, -x_n + \frac{b_n}{y_n} + \frac{a_n}{y^2_n} \right).\label{varphimap}
\end{equation}
We also introduce the variables $s_n = 1 / x_n$ and $t_n = 1 / y_n$ and we cover $\mathbb{P}^1 \times \mathbb{P}^1$ with four copies of $\mathbb{C}^2$, as:
\begin{equation}
\mathbb{P}^1 \times \mathbb{P}^1 = (x_n, y_n) \cup (x_n, t_n) \cup (s_n, y_n) \cup (s_n, t_n).\label{p1p1cover}
\end{equation}
Obviously, the point $(s_n, y_n) = (0, 0)$ is an indeterminate point for the map $\varphi_n$ which we shall try to regularise by a `blow-up', i.e. by introducing two new coordinate charts that will replace the single point $(s_n, y_n) = (0, 0)$ (which will be called the base-point for the blow-up):
\begin{equation}
(s_n, y_n) \leftarrow \left( s_n, \frac{y_n}{s_n} \right) \cup \left( \frac{s_n}{y_n}, y_n \right).
\end{equation}
The map can now be expressed as
\begin{equation}
y_{n+1} = \frac{1}{y^2_n \frac{s_n}{y_n}} \left( - y_n + b_n y_n \frac{s_n}{y_n} + a_n \frac{s_n}{y_n} \right),\label{varphimap1}
\end{equation}
which, however, is now indeterminate at $\left( \frac{s_n}{y_n}, y_n \right) = (0, 0)$ in the second new coordinate chart.
Hence we introduce two new coordinate charts
\begin{equation}
\left( \frac{s_n}{y_n}, y_n \right) \leftarrow \left( \frac{s_n}{y_n}, \frac{y^2_n}{s_n} \right) \cup \left( \frac{s_n}{y^2_n}, y_n \right),\label{bu2}
\end{equation}
at the base-point $\left( \frac{s_n}{y_n}, y_n \right) = (0, 0)$, after which the mapping is given by
\begin{equation}
y_{n+1} = \frac{1}{y^2_n \frac{s_n}{y^2_n}} \left( - 1 + b_n y_n \frac{s_n}{y^2_n} + a_n \frac{s_n}{y^2_n} \right),
\end{equation}
in terms of the second coordinate chart in \eqref{bu2}. However, this does not yet resolve the indeterminacy, which now manifests itself at $\left( \frac{s_n}{y^2_n}, y_n \right) = \left( \frac{1}{a_n}, 0 \right)$, which also requires blowing-up. Note that this base-point now depends on the independent variable $n$ through the parameter function $a_n$, whereas the base-points in the previous two blow-ups were independent of $n$.
This third blow-up is carried out by means of the coordinate charts
\begin{equation}
\left( \frac{s_n}{y^2_n} - \frac{1}{a_n}, y_n \right) \leftarrow \left( \frac{s_n}{y^2_n} - \frac{1}{a_n}, \frac{y_n}{\frac{s_n}{y^2_n} - \frac{1}{a_n}} \right) \cup \left( \frac{1}{y_n} \left( \frac{s_n}{y^2_n} - \frac{1}{a_n} \right), y_n \right),
\end{equation}
in terms of which we obtain
\begin{equation}
	y_{n+1} = \frac{1}{y_n \frac{s_n}{y^2_n}} \left( \frac{a_n}{y_n} \left( \frac{s_n}{y^2_n} - \frac{1}{a_n} \right) + b_n \frac{s_n}{y^2_n} \right),
\end{equation}
which still has an indeterminacy at the point $\left( \frac{1}{y_n} \left( \frac{s_n}{y^2_n} - \frac{1}{a_n} \right), y_n \right) = \left( -\frac{b_n}{a^2_n}, 0 \right)$. Note that this point now also involves the second parameter, $b_n$. A blow-up at this base-point, using the coordinate charts
\begin{multline}
\left( \frac{1}{y_n} \left( \frac{s_n}{y^2_n} - \frac{1}{a_n} \right) + \frac{b_n}{a^2_n}, y_n \right) \leftarrow \left( \frac{1}{y_n} \left( \frac{s_n}{y^2_n} - \frac{1}{a_n} \right) + \frac{b_n}{a^2_n}, \frac{y_n}{\frac{1}{y_n} \left( \frac{s_n}{y^2_n} - \frac{1}{a_n} \right) + \frac{b_n}{a^2_n}} \right) \\
\cup \left( \frac{1}{y_n} \left( \frac{1}{y_n} \left( \frac{s_n}{y^2_n} - \frac{1}{a_n} \right) + \frac{b_n}{a^2_n} \right), y_n \right),\hskip2.5cm
\end{multline}
finally lifts the indeterminacy, as one obtains:
\begin{equation}
	y_{n+1} = \frac{1}{\frac{s_n}{y^2_n}} \left( \frac{1}{y_n} \left( \frac{a_n}{y_n} \left( \frac{s_n}{y^2_n} - \frac{1}{a_n} \right) + \frac{b_n}{a^2_n} \right) + \frac{b_n}{a^2_n y_n} \left( \frac{s_n}{y^2_n} - \frac{1}{a_n} \right) \right).
\end{equation}

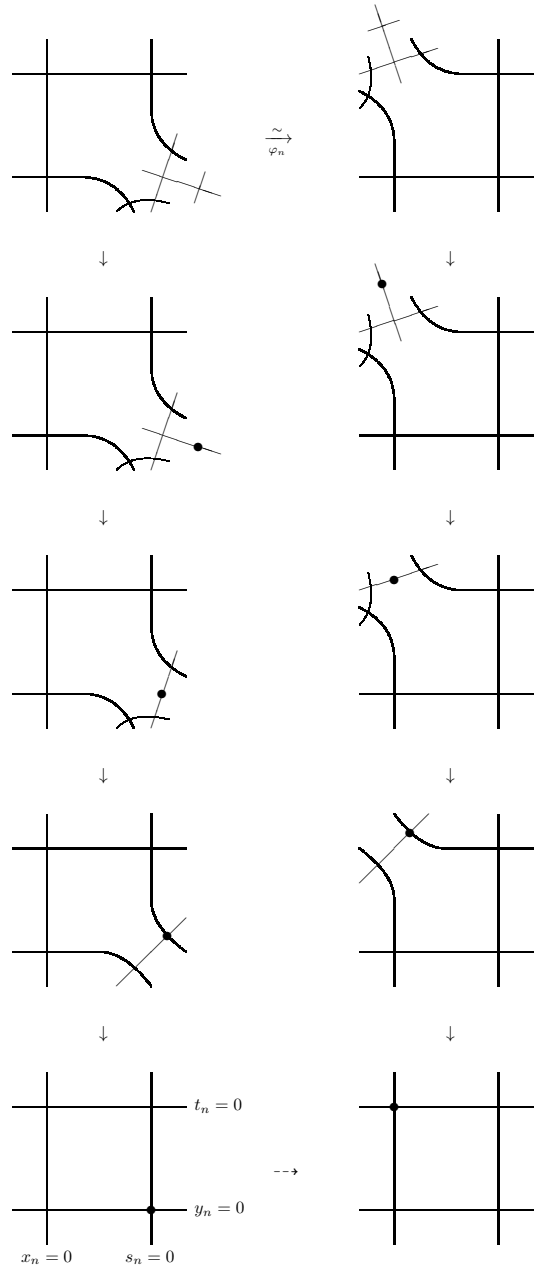
\begin{figure}[!h]
\begin{center}
\resizebox{7cm}{!}{
\begin{picture}(300, 740)
	{\thicklines
	\put(0, 30){\line(1, 0){100}}
	\put(0, 90){\line(1, 0){100}}
	\put(20, 10){\line(0, 1){100}}
	\put(80, 10){\line(0, 1){100}}
	}
	
	\put(80, 30){\circle*{5}}

	\put(5, 0){$x_n = 0$}
	\put(65, 0){$s_n = 0$}
	\put(105, 28){$y_n = 0$}
	\put(105, 88){$t_n = 0$}

	\put(50, 130){$\downarrow$}

	{\thicklines
	\put(0, 180){\line(1, 0){50}}
	\put(0, 240){\line(1, 0){100}}
	\put(20, 160){\line(0, 1){100}}
	\put(80, 210){\line(0, 1){50}}
	\qbezier(80, 210)(80, 195)(100, 180)
	\qbezier(50, 180)(65, 180)(80, 160)
	}
	\put(60, 160){\line(1, 1){40}}
	
	\put(89, 189){\circle*{5}}

	\put(50, 280){$\downarrow$}

	{\thicklines
	\put(0, 330){\line(1, 0){40}}
	\put(0, 390){\line(1, 0){100}}
	\put(20, 310){\line(0, 1){100}}
	\put(80, 370){\line(0, 1){40}}
	\qbezier(80, 370)(80, 350)(100, 340)
	\qbezier(40, 330)(60, 330)(70, 310)
	}
	\qbezier(60, 310)(70, 320)(90, 315)
	\put(80, 310){\line(1, 3){15}}
	
	\put(86, 330){\circle*{5}}

	\put(50, 430){$\downarrow$}

	{\thicklines
	\put(0, 480){\line(1, 0){40}}
	\put(0, 540){\line(1, 0){100}}
	\put(20, 460){\line(0, 1){100}}
	\put(80, 520){\line(0, 1){40}}
	\qbezier(80, 520)(80, 500)(100, 490)
	\qbezier(40, 480)(60, 480)(70, 460)
	}
	\qbezier(60, 460)(70, 470)(90, 465)
	\put(80, 460){\line(1, 3){15}}
	\put(75, 484){\line(3, -1){45}}
	
	\put(107, 473){\circle*{5}}

	\put(50, 580){$\downarrow$}

	{\thicklines
	\put(0, 630){\line(1, 0){40}}
	\put(0, 690){\line(1, 0){100}}
	\put(20, 610){\line(0, 1){100}}
	\put(80, 670){\line(0, 1){40}}
	\qbezier(80, 670)(80, 650)(100, 640)
	\qbezier(40, 630)(60, 630)(70, 610)
	}
	\qbezier(60, 610)(70, 620)(90, 615)
	\put(80, 610){\line(1, 3){15}}
	\put(75, 634){\line(3, -1){45}}
	
	\put(105, 615){\line(1, 3){6}}

	{\thicklines
	\put(200, 30){\line(1, 0){100}}
	\put(200, 90){\line(1, 0){100}}
	\put(220, 10){\line(0, 1){100}}
	\put(280, 10){\line(0, 1){100}}
	}
	
	\put(220, 90){\circle*{5}}

	\put(250, 130){$\downarrow$}

	{\thicklines
	\put(220, 160){\line(0, 1){50}}
	\put(280, 160){\line(0, 1){100}}
	\put(200, 180){\line(1, 0){100}}
	\put(250, 240){\line(1, 0){50}}
	\qbezier(220, 210)(220, 225)(200, 240)
	\qbezier(250, 240)(235, 240)(220, 260)
	}
	\put(200, 220){\line(1, 1){40}}
	
	\put(229, 249){\circle*{5}}

	\put(250, 280){$\downarrow$}

	{\thicklines
	\put(220, 310){\line(0, 1){40}}
	\put(280, 310){\line(0, 1){100}}
	\put(200, 330){\line(1, 0){100}}
	\put(260, 390){\line(1, 0){40}}
	\qbezier(220, 350)(220, 370)(200, 380)
	\qbezier(260, 390)(240, 390)(230, 410)
	}
	\qbezier(200, 370)(210, 380)(205, 400)
	\put(200, 390){\line(3, 1){45}}
	
	\put(220, 396){\circle*{5}}

	\put(250, 430){$\downarrow$}

	{\thicklines
	\put(220, 460){\line(0, 1){40}}
	\put(280, 460){\line(0, 1){100}}
	\put(200, 480){\line(1, 0){100}}
	\put(260, 540){\line(1, 0){40}}
	\qbezier(220, 500)(220, 520)(200, 530)
	\qbezier(260, 540)(240, 540)(230, 560)
	}
	\qbezier(200, 520)(210, 530)(205, 550)
	\put(200, 540){\line(3, 1){45}}
	\put(224, 535){\line(-1, 3){15}}
	
	\put(213, 568){\circle*{5}}

	\put(250, 580){$\downarrow$}

	{\thicklines
	\put(220, 610){\line(0, 1){40}}
	\put(280, 610){\line(0, 1){100}}
	\put(200, 630){\line(1, 0){100}}
	\put(260, 690){\line(1, 0){40}}
	\qbezier(220, 650)(220, 670)(200, 680)
	\qbezier(260, 690)(240, 690)(230, 710)
	}
	\qbezier(200, 670)(210, 680)(205, 700)
	\put(200, 690){\line(3, 1){45}}
	\put(224, 685){\line(-1, 3){15}}
	
	\put(205, 715){\line(3, 1){18}}

	\put(150, 50){$\dashrightarrow$}
	\put(145, 650){$\xrightarrow[\varphi_n]{\sim}$}

\end{picture}
}
\end{center}\vskip-.0cm
\caption{Diagram showing, on the left, the base-points in the successive blow-ups of the map $\varphi_n$ and the exceptional curves resulting from these blow-ups. Same on the right for the indeterminate points of the inverse map $\varphi^{-1}_n$.}
\label{fig1}\vskip-.0cm
\end{figure}

This sequence of four succesive blow-ups at the point $(s_n, y_n) = (0, 0)$ is depicted on the left in Figure \ref{fig1}, together with the resulting configuration of exceptional curves. These curves can be thought of as (local) copies of $\mathbb{P}^1$, spliced in as it were, at the base-point of each blow-up and that are described by the coordinate charts introduced for that blow-up.

However, this does not regularise all indeterminate points for the mapping $\varphi_n$. The easiest way to resolve the remaining indeterminacies and lift $\varphi_n$ to a well-defined automorphism, is to consider its inverse map
\begin{equation}
	\varphi^{-1}_n \colon\quad \mathbb{P}^1 \times \mathbb{P}^1 \dashrightarrow \mathbb{P}^1 \times \mathbb{P}^1,
	\qquad (x_{n+1}, y_{n+1}) \mapsto (x_n, y_n) = \left( -y_{n+1} + \frac{b_n}{x_{n+1}} + \frac{a_n}{x^2_{n+1}}, x_{n+1} \right),
\end{equation}
which is undefined at $(x_{n+1}, t_{n+1}) = (0, 0)$. As was the case for $\varphi_n$, this indeterminacy can be resolved by four succesive blow-ups, the base-points and coordinate charts of which are given below (the resulting mappings are however omitted, for brevity).
\begin{gather}
\text{blow-up at ~} (x_{n+1}, t_{n+1}) = (0, 0)~:\quad (x_{n+1}, t_{n+1}) \leftarrow \left( x_{n+1}, \frac{t_{n+1}}{x_{n+1}} \right) \cup \left( \frac{x_{n+1}}{t_{n+1}}, t_{n+1} \right)\\\nonumber\\
\text{blow-up at ~}\left( x_{n+1}, \frac{t_{n+1}}{x_{n+1}} \right) = (0, 0)~:\quad \left( x_{n+1}, \frac{t_{n+1}}{x_{n+1}} \right) \leftarrow \left( x_{n+1}, \frac{t_{n+1}}{x^2_{n+1}} \right) \cup \left( \frac{x^2_{n+1}}{t_{n+1}}, \frac{t_{n+1}}{x_{n+1}} \right)\\\nonumber\\
\text{blow-up at ~}\left( x_{n+1}, \frac{t_{n+1}}{x^2_{n+1}} \right) = \left( 0, \frac{1}{a_n} \right)~:\quad \left( x_{n+1}, \frac{t_{n+1}}{x^2_{n+1}} - \frac{1}{a_n} \right) \leftarrow \left( x_{n+1}, \frac{1}{x_{n+1}}\left( \frac{t_{n+1}}{x^2_{n+1}} - \frac{1}{a_n} \right) \right)\nonumber\\ 
\hskip9cm\cup \left( \frac{x_{n+1}}{\frac{t_{n+1}}{x^2_{n+1}} - \frac{1}{a_n}}, \frac{t_{n+1}}{x^2_{n+1}} - \frac{1}{a_n} \right)\\
\text{blow-up at~}\left( x_{n+1}, \frac{1}{x_{n+1}}\left( \frac{t_{n+1}}{x^2_{n+1}} - \frac{1}{a_n} \right) \right) = \left( 0, -\frac{b_n}{a^2_n} \right)~:\hskip6.5cm\nonumber\\ \left( x_{n+1}, \frac{1}{x_{n+1}}\left( \frac{t_{n+1}}{x^2_{n+1}} - \frac{1}{a_n} \right) + \frac{b_n}{a^2_n} \right) \leftarrow \left(x_{n+1}, \frac{1}{x_{n+1}}\left(\frac{1}{x_{n+1}}\left( \frac{t_{n+1}}{x^2_{n+1}} - \frac{1}{a_n} \right)+ \frac{b_n}{a^2_n}\right)\right)\nonumber\\
\hskip5cm\cup \left( \frac{x_{n+1}}{ \frac{1}{x_{n+1}}\left( \frac{t_{n+1}}{x^2_{n+1}} - \frac{1}{a_n} \right) + \frac{b_n}{a^2_n}}, \frac{1}{x_{n+1}}\left( \frac{t_{n+1}}{x^2_{n+1}} - \frac{1}{a_n} \right) + \frac{b_n}{a^2_n} \right).
\end{gather}
These blow-ups and their corresponding exceptional curves are shown on the right in Figure \ref{fig1}. Note that, here as well, the base-points in the last two blow-ups depend on $n$, through the parameters $a_n$ and $b_n$. Moreover, it is easily verified that the full set of eight blow-ups described above resolves all indeterminacies in the autonomous case of \eqref{varphimap}, i.e. when the parameters $a_n$ and $b_n$ are constant for all $n$. In the non-autonomous case however, these parameters have to satisfy certain constraints for the blow-ups to fully resolve the indeterminacies in the mapping. Indeed, it is easy to check that $\varphi_n$ maps the point $(x_n, t_n) = (0, 0)$, i.e. the first indeterminate point for $\varphi^{-1}_{n-1}$, into the first indeterminate point $(s_{n+1}, y_{n+1}) = (0, 0)$ for $\varphi_{n+1}$. Similarly, $\varphi_n$ maps the second indeterminate point $\left( x_n, \frac{t_n}{x_n} \right) = (0, 0)$ for $\varphi^{-1}_{n-1}$ into the second indeterminate point $\left( \frac{s_{n+1}}{y_{n+1}}, y_{n+1} \right) = (0, 0)$ of $\varphi_{n+1}$ (in its \eqref{varphimap1} avatar).

In order to analyse the relations between the other indeterminate points, we define:\begin{gather}
P_n\,: ~\left( x_n, \frac{t_n}{x^2_n} \right) = \left(0, \frac{1}{a_{n-1}}\right)\,,\qquad
Q_{n+1}\,: ~\left( \frac{s_{n+1}}{y^2_{n+1}}, y_{n+1} \right) = \left( \frac{1}{a_{n+1}}, 0 \right)\,,\\
\overline{P_n}\,:~\left( \frac{s_{n+1}}{y^2_{n+1}}, y_{n+1} \right) = \left( \frac{1}{a_{n-1}}, 0 \right),
\end{gather}
and
\begin{gather}
R_n\,: ~\left( x_n, \frac{1}{x_n}\left( \frac{t_n}{x^2_n} - \frac{1}{a_{n-1}} \right) \right) = \left( 0, -\frac{b_{n-1}}{a^2_{n-1}} \right)\,,\\
S_{n+1}\,:~ \left( \frac{1}{y_{n+1}}\left( \frac{s_{n+1}}{y^2_{n+1}} - \frac{1}{a_{n+1}} \right), y_{n+1} \right) = \left( -\frac{b_{n+1}}{a^2_{n+1}}, 0 \right)\,.\\\
\overline{R_n}\,: ~\left( \frac{1}{y_{n+1}}\left( \frac{s_{n+1}}{y^2_{n+1}} - \frac{1}{a_{n+1}} \right), y_{n+1} \right) = \left( \frac{-2 b_n + b_{n-1}}{a^2_{n-1}}, 0 \right).
\end{gather}

\begin{figure}[!h]
\begin{center}
\resizebox{8.5cm}{!}{
\begin{picture}(300, 130)
	{\thicklines
	\put(0, 30){\line(1, 0){40}}
	\put(80, 70){\line(0, 1){40}}
	\qbezier(80, 70)(80, 50)(100, 40)
	\qbezier(40, 30)(60, 30)(70, 10)
	}
	\qbezier(60, 10)(70, 20)(90, 15)
	\put(80, 10){\line(1, 3){15}}
	\put(75, 34){\line(3, -1){45}}
	
	\put(105, 15){\line(1, 3){6}}

	{\thicklines
	\put(20, 10){\line(0, 1){40}}
	\put(60, 90){\line(1, 0){40}}
	\qbezier(20, 50)(20, 70)(0, 80)
	\qbezier(60, 90)(40, 90)(30, 110)
	}
	\qbezier(0, 70)(10, 80)(5, 100)
	\put(0, 90){\line(3, 1){45}}
	
	\put(20, 96){\circle*{5}}

	\put(20, 86){$P_n$}

	{\thicklines
	\put(220, 10){\line(0, 1){40}}
	\put(260, 90){\line(1, 0){40}}
	\qbezier(220, 50)(220, 70)(200, 80)
	\qbezier(260, 90)(240, 90)(230, 110)
	}
	\qbezier(200, 70)(210, 80)(205, 100)
	\put(200, 90){\line(3, 1){45}}
	\put(224, 85){\line(-1, 3){15}}
	
	\put(205, 115){\line(3, 1){18}}

	{\thicklines
	\put(200, 30){\line(1, 0){40}}
	\put(280, 70){\line(0, 1){40}}
	\qbezier(280, 70)(280, 50)(300, 40)
	\qbezier(240, 30)(260, 30)(270, 10)
	}
	\qbezier(260, 10)(270, 20)(290, 15)
	\put(280, 10){\line(1, 3){15}}
	
	\put(286, 30){\circle*{5}}
	
	\put(290, 25){$Q_{n+1}$}

	\put(145, 50){$\xrightarrow[\varphi_n]{\sim}$}

\end{picture}
}
\end{center}\vskip-.0cm
\caption{The link between the base-point $P_n$ for the third blow-up for $\varphi_{n-1}^{-1}$ and the base-point $Q_{n+1}$ for the third blow-up for $\varphi_{n+1}$.}
\label{fig2}\vskip-.0cm
\end{figure}
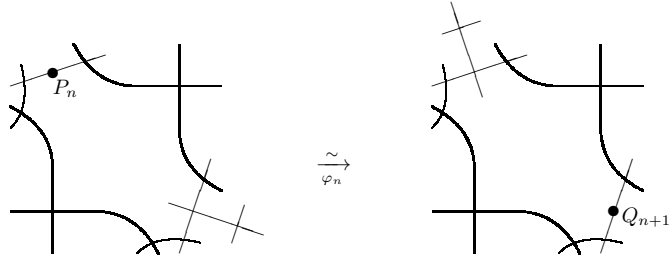

It is easily verified that $\overline{P_n}$ is the image of $P_n$ under the mapping $\varphi_n$. Moreover, in the autonomous case ($a_n=a, b_n=b, ^\forall\! n$) it is clear that  $\overline{P_n} = Q_{n+1}$, as depicted in Figure \ref{fig2}. 

\begin{figure}[h!]
\begin{center}
\resizebox{9cm}{!}{
\begin{picture}(300, 130)
	{\thicklines
	\put(0, 30){\line(1, 0){40}}
	\put(80, 70){\line(0, 1){40}}
	\qbezier(80, 70)(80, 50)(100, 40)
	\qbezier(40, 30)(60, 30)(70, 10)
	}
	\qbezier(60, 10)(70, 20)(90, 15)
	\put(80, 10){\line(1, 3){15}}
	\put(75, 34){\line(3, -1){45}}
	
	\put(105, 15){\line(1, 3){6}}

	{\thicklines
	\put(20, 10){\line(0, 1){40}}
	\put(60, 90){\line(1, 0){40}}
	\qbezier(20, 50)(20, 70)(0, 80)
	\qbezier(60, 90)(40, 90)(30, 110)
	}
	\qbezier(0, 70)(10, 80)(5, 100)
	\put(0, 90){\line(3, 1){45}}
	\put(24, 85){\line(-1, 3){15}}
	
	\put(13, 118){\circle*{5}}
	
	\put(16, 118){$R_n$}

	{\thicklines
	\put(220, 10){\line(0, 1){40}}
	\put(260, 90){\line(1, 0){40}}
	\qbezier(220, 50)(220, 70)(200, 80)
	\qbezier(260, 90)(240, 90)(230, 110)
	}
	\qbezier(200, 70)(210, 80)(205, 100)
	\put(200, 90){\line(3, 1){45}}
	\put(224, 85){\line(-1, 3){15}}
	
	\put(205, 115){\line(3, 1){18}}

	{\thicklines
	\put(200, 30){\line(1, 0){40}}
	\put(280, 70){\line(0, 1){40}}
	\qbezier(280, 70)(280, 50)(300, 40)
	\qbezier(240, 30)(260, 30)(270, 10)
	}
	\qbezier(260, 10)(270, 20)(290, 15)
	\put(280, 10){\line(1, 3){15}}
	\put(275, 34){\line(3, -1){45}}
	
	\put(307, 23){\circle*{5}}
	
	\put(307, 27){$S_{n+1}$}

	\put(145, 50){$\xrightarrow[\varphi_n]{\sim}$}

\end{picture}
}
\end{center}\vskip-.0cm
\caption{The link between the base-point $R_n$ for the fourth blow-up for $\varphi_{n-1}^{-1}$ and the base-point $S_{n+1}$ for the fourth blow-up for $\varphi_{n+1}$.}
\label{fig3}\vskip-.0cm
\end{figure}
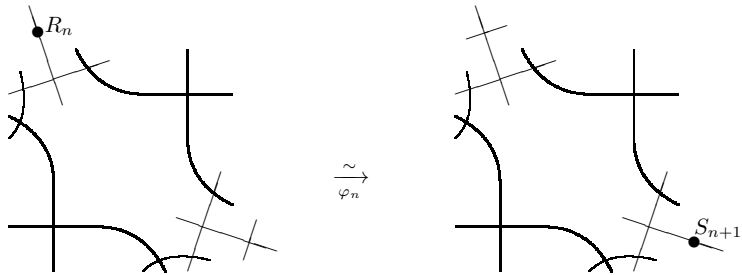

The same is true for the points $\overline{R_n} \equiv \varphi_n(R_n)$ and $S_{n+1}$ which automatically coincide in the autonomous case (cf. Figure \ref{fig3}). Hence, in the autonomous case, since $\varphi_n$ maps the indeterminate points of $\varphi_{n-1}^{-1}$ to those of $\varphi_{n+1}$ (and vice versa, mutatis mutandis) and since all these indeterminacies have been resolved, in general, by blowing up $\mathbb{P}^1 \times \mathbb{P}^1$ eight times at every value of $n$ , the mapping $\varphi_n$ has been successfully regularised. Now, it is clear that these same eight blow-ups of $\mathbb{P}^1 \times \mathbb{P}^1$ can also fully regularise the non-autonomous mapping if one imposes certain constraints on the parameter functions. In particular, $\overline{P_n}$ and $Q_{n+1}$ will coincide if and only if $a_{n+1} = a_{n-1}$ holds, which is nothing but the first constraint obtained from singularity confinement for the mapping \eqref{eqiii}. Furthermore, under this constraint we can also require $\overline{R_n} = S_{n+1}$, which amounts to requiring $b_n$ to satisfy $b_{n+1} - 2 b_n + b_{n-1} = 0$, exactly as in the singularity confinement analysis.

It is also clear that taking $b_n=0 \,(^\forall\!n)$ in no way affects the number of blow-ups needed to fully regularise the mapping, nor does it influence the configuration of exceptional curves obtained from these blow-ups. This statement is in fact equivalent to saying that the presence or absence of the term $b_n/x_n$ in the right-hand side of the mapping \eqref{eqiii} does not alter its singularity pattern, which is the fundamental observation that underlies the full-deautonomisation procedure. To explain the precise link between the singularity pattern and the blow-ups carried out above, it is important however to take a closer look at the configuration of exceptional curves that is obtained from them. 

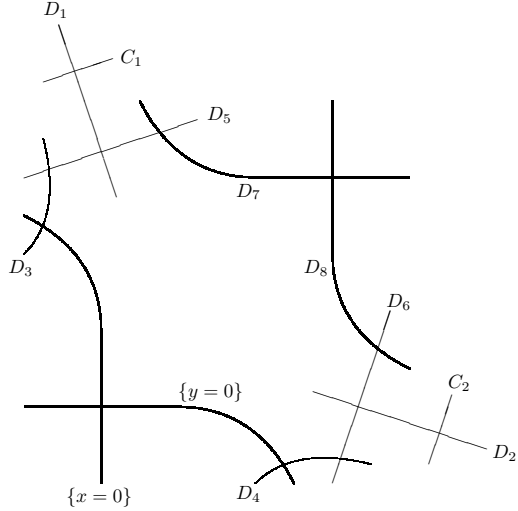
\begin{figure}[h!]
\begin{center}
\resizebox{6cm}{!}{
\begin{picture}(230, 265)
	
	{\thicklines
	\put(40, 20){\line(0, 1){80}}
	\put(120, 180){\line(1, 0){80}}
	\qbezier(40, 100)(40, 140)(0, 160)
	\qbezier(120, 180)(80, 180)(60, 220)
	}
	\qbezier(0, 140)(20, 160)(10, 200)
	\put(0, 180){\line(3, 1){90}}
	\put(48, 170){\line(-1, 3){30}}
	
	\put(10, 230){\line(3, 1){36}}

	{\thicklines
	\put(0, 60){\line(1, 0){80}}
	\put(160, 140){\line(0, 1){80}}
	\qbezier(160, 140)(160, 100)(200, 80)
	\qbezier(80, 60)(120, 60)(140, 20)
	}
	\qbezier(120, 20)(140, 40)(180, 30)
	\put(160, 20){\line(1, 3){30}}
	\put(150, 68){\line(3, -1){90}}
	
	\put(210, 30){\line(1, 3){12}}

	\put(10, 265){$D_1$}
	\put(243, 33){$D_2$}
	\put(-8, 130){$D_3$}
	\put(110, 12){$D_4$}
	\put(95, 210){$D_5$}
	\put(188, 112){$D_6$}
	\put(110, 170){$D_7$}
	\put(145, 130){$D_8$}
	
	\put(50, 240){$C_1$}
	\put(220, 70){$C_2$}
	
	\put(22, 10){$\{ x = 0 \}$}
	\put(80, 65){$\{ y = 0 \}$}
	
\end{picture}
}
\end{center}\vskip-.3cm
\caption{Diagrammatic representation of the family of surfaces on which $\varphi_n$ acts as an automorphism. The labels $\{x=0\}$ and  $\{y=0\}$ refer to the strict transforms of the corresponding curves in  $\mathbb{P}^1 \times \mathbb{P}^1$.}
\label{fig4}\vskip-.0cm
\end{figure}

The sequence of blow-ups of $\mathbb{P}^1 \times \mathbb{P}^1$ described above, produces a rational surface \cite{hart} that contains all the exceptional curves obtained from the blow-ups (these consist of all the points that correspond to the base-points of the blow-ups) and which is obtained by glueing together all the different coordinate charts used in the blow-ups. In fact, to be completely accurate, one should say {\sl a family} of surfaces, since the base-points for four of the blow-ups are $n$-dependent and the corresponding surfaces will therefore not be strictly identical. However, algebraically speaking, only the relative positions and, especially, the intersection pattern of these curves matters and one can therefore represent this entire family of surfaces by a single diagram, such as in Figure \ref{fig4}.
The curves depicted in this figure are part of a free Abelian group, the so called Picard group \cite{hart}, which is in this case of rank 10 as it is obtained by 8 blow-ups from $\mathbb{P}^1 \times \mathbb{P}^1$. In fact, we can take the curves $D_1, D_2, \hdots, D_8, C_1$ and $C_2$, as a basis that will generate the entire group. Note that the curves labelled with $D$ differ in nature from the curves $C_1$ and $C_2$ as the latter have self-intersection -1 whereas the former have self-intersection -2 and, hence, can not be trivially `blown-down'. In fact, the intersection pattern of the curves with self-intersection -2 has the form of a Dynkin diagram of type $D^{(1)}_7$. This type of characterization of the surface plays an important role in the classification of discrete Painlev\'e equations obtained by Sakai \cite{sakai}, where the non-autonomous mapping \eqref{eqiii} that satisfies the two conditions $a_{n+2}=a_n$ and $b_{n+2} - 2 b_{n+1} + b_n=0$ is classified as a discrete Painlev\'e with symmetries associated with the affine Weyl-type group $A^{(1)}_{1, |\alpha|^2 = 4}$.

Moreover, as is clear from the correspondence between the indeterminate points of $\varphi_n^{-1}$ and $\varphi_{n+1}$ under the mapping $\varphi_n$, explained above, $\varphi_n$ induces the following automorphism on the Picard group:
\begin{gather}
D_1 \to D_2 \to D_3 \to D_4 \to D_1\,,\qquad D_5 \to D_6 \to D_5\,,\qquad D_7 \to D_8 \to D_7\label{Dseq1}\\
\{ y = 0 \} \to C_1 \to C_2 \to \{ x = 0 \}\,,\label{yseq1}
\end{gather}
where $\{ x = 0 \}$ is found to be linearly equivalent to 
\begin{equation}
- D_1 + 2 D_2 - D_3 + D_4 - D_5 + 2 D_6 + D_8 - C_1 + 2C_2\,.
\end{equation}
Note that the sequence of values for $y$ in \eqref{yseq1} is nothing but the singularity pattern $\{ 0, \infty^2, 0\}$.

With respect to the basis for the Picard group we have chosen, this automorphism, which we shall denote by $\varphi_{*}$, can be written in matrix form as:

\begin{equation}
\left(\begin{array}{cccccccc|crc}
	0 & 0 & 0 & 1 &   &   &   &   & 0 & -1  \\
	1 & 0 & 0 & 0 &   &   &   &   & 0 & 2 \\
	0 & 1 & 0 & 0 &   &   &   &   & 0 & -1  \\
	0 & 0 & 1 & 0 &   &   &   &   & 0 & 1 \\
	  &   &   &   & 0 & 1 &   &   & 0 & -1  \\
	  &   &   &   & 1 & 0 &   &   & 0 & 2 \\
	  &   &   &   &   &   & 0 & 1 & 0 & 0  \\
	  &   &   &   &   &   & 1 & 0 & 0 & 1 \\ \hline
	  &   &   &   &   &   &   &   &  &    \\
	  & & & \mbox{{\LARGE $0$}}  & & & & &~ \mbox{\hspace{1em}{\LARGE $A$}}& \\
	  &  & &  & &   &   &   &  &   
	  \end{array}\right)\!,
\end{equation}
where $A=\begin{pmatrix}
	0	&	-1 \\
	1	&	2
\end{pmatrix}
$.
It is well-known \cite{take} that the algebraic entropy ${\mathcal E}$ of a birational map such as \eqref{varphimap} can be obtained from the largest eigenvalue, $\lambda_*$, for the induced automorphism $\varphi_{*}$:
\begin{equation}
{\mathcal E} := \lim_{m\rightarrow\infty}\frac{1}{m}\log\big({\rm deg}(\varphi_{n+m-1}\varphi_{n+m-2}\cdots\varphi_n)\big) = \log|\lambda_*|,\label{algent}
\end{equation}
where by ${\rm deg(\varphi)}$, we mean the maximum of the degrees of the numerator and denominator of the birational map $\varphi$. Of course, as is clear from the above matrix, its upper-left sub-matrix (corresponding to the sequences \eqref{Dseq1} of $D$'s) cannot yield an eigenvalue with modulus greater than 1 since it is unitary. Hence, the part of the matrix that decides on the positivity of the entropy, i.e. on the integrability or non-integrability of the mapping $\varphi_n$, is the sub-matrix $A$ which has
\begin{equation}
	\det(\lambda I - A) = \lambda^2 - 2 \lambda + 1,
\end{equation}
as characteristic polynomial. All eigenvalues of $\varphi_*$ therefore have modulus 1 and the algebraic entropy of $\varphi_n$ is zero. The mapping, subject to the conditions $a_{n+2}=a_n$ and $b_{n+2} - 2 b_{n+1} + b_n=0$ which are needed in the above regularisation, is therefore integrable. It is important to note that the characteristic polynomial for the $A$ part of $\varphi_*$ is identical to that for the only non-trivial constraint on the parameters in the mapping $\varphi_n$: $b_{n+2} - 2 b_{n+1} + b_n=0$ (as $a_n$ such that $a_{n+2}=a_n$, corresponds to a simple gauge transformation and, as such, has to be considered as a trivial constraint vis-\`a-vis the integrability of the mapping). We shall see on a second example that similar conclusions can be drawn in the case of non-integrable mappings.

\subsection{The case of a non-integrable mapping}
Next, we consider the special case of mapping \eqref{eqv} when $c_n=d_n=0$. The analysis that follows can of course also be carried out for the general case, i.e. for non-zero $c_n$ and $d_n$, but besides complicating the relevant mathematical expressions to the point where they become almost illegible, such a generalization does not alter the conclusions we shall obtain in the restricted case.

Here as well, we interpret the mapping as a birational map on $\mathbb{P}^1 \times \mathbb{P}^1$:
\begin{equation}
	\phi_n \colon\quad \mathbb{P}^1 \times \mathbb{P}^1 \dashrightarrow \mathbb{P}^1 \times \mathbb{P}^1,
	\qquad (x_n, y_n) \mapsto (x_{n+1}, y_{n+1}) = \left( y_n, - x_n + \frac{b_n}{y_n} + \frac{a_n}{y^4_n} \right),\label{phimap}
\end{equation}
which is undefined at $(s_n, y_n) = (0, 0)$ (in the coordinates introduced in \eqref{p1p1cover}) and eight successive blow-ups are needed to lift this indeterminacy. However, as their precise mathematical expressions become quite involved, we prefer to give these 8 base-points $Q_n^{(i)}$ ($i=1\cdots,8$), with the corresponding coordinate charts and the maps obtained after each blow-up, in an appendix to this paper.

As in the previous case, the remaining indeterminacies are best resolved at the level of the inverse map
\begin{equation}
	\phi^{-1}_n \colon\quad \mathbb{P}^1 \times \mathbb{P}^1 \dashrightarrow \mathbb{P}^1 \times \mathbb{P}^1,
	\qquad (x_{n+1}, y_{n+1}) \mapsto (x_n, y_n) = \left( -y_{n+1} + \frac{b_n}{x_{n+1}} + \frac{a_n}{x^4_{n+1}}, x_{n+1} \right),
\end{equation}
which is indeterminate at the point $(x_{n+1}, t_{n+1}) = (0, 0)$. Again, eight blow-ups are required to lift the indeterminacy and the base-points $P_{n+1}^{(i)}$ ($i=1\cdots,8$) and coordinate charts for these blow-ups are also given in the appendix.

Now, it can be easily checked that $\phi_n$ maps the base-points $P_n$ of the first four blow-ups for $\phi_{n-1}^{-1}$ into the base-points $Q_{n+1}$ of the corresponding blow-ups for $\phi_{n+1}$, i.e.: $\phi_n\left(P_n^{(i)}\right)= Q_{n+1}^{(i)}$ for $i=1, 2, 3$ and $4$. For the base-points of the fifth blow-up we find that
\begin{equation}
\phi\left(P_n^{(5)}\right): ~\left( \frac{s_{n+1}}{y^4_{n+1}}, y_{n+1} \right) = \left( \frac{1}{a_{n-1}}, 0 \right)
\end{equation}
matches up automatically with the base-point $Q_{n+1}^{(5)} : \left( \frac{s_{n+1}}{y^4_{n+1}}, y_{n+1} \right) = \left( \frac{1}{a_{n+1}}, 0 \right)$ for the fifth blow-up of $\phi_{n+1}$, in the autonomous case, but that the parameter $a_n$ needs to satisfy the constraint
\begin{equation}
a_{n+1} = a_{n-1},\label{phicond1}
\end{equation}
for this match to occur in general, in the non-autonomous case.
Under this constraint, the images under $\phi_n$ of the next two base-points for  $\phi_{n-1}^{-1}$ again coincide naturally with those for $\phi_{n+1}$, i.e.: $\phi_n\left(P_n^{(i)}\right)= Q_{n+1}^{(i)}$, for $i=6$ and $7$. However, the remaining base-points only match up automatically in the autonomous case. For the general non-autonomous mapping, one finds that
\begin{equation}
\phi_n\left(P_n^{(8)}\right): \left( \frac{1}{y^3_{n+1}} \left( \frac{s_{n+1}}{y^4_{n+1}} - \frac{1}{a_{n+1}} \right), y_{n+1} \right) = \left( \frac{- 4 b_n + b_{n-1}}{a^2_{n-1}}, 0 \right),
\end{equation}
will coincide with
\begin{equation}
Q_{n+1}^{(8)}: \left( \frac{1}{y^3_{n+1}} \left( \frac{s_{n+1}}{y^4_{n+1}} - \frac{1}{a_{n+1}} \right), y_{n+1} \right) = \left( - \frac{b_{n+1}}{a^2_{n+1}}, 0 \right),
\end{equation}
if and only if, under the constraint \eqref{phicond1},  $b_n$ satisfies the condition
\begin{equation}
b_{n+1} - 4 b_n + b_{n-1} = 0.\label{phicond2}
\end{equation}

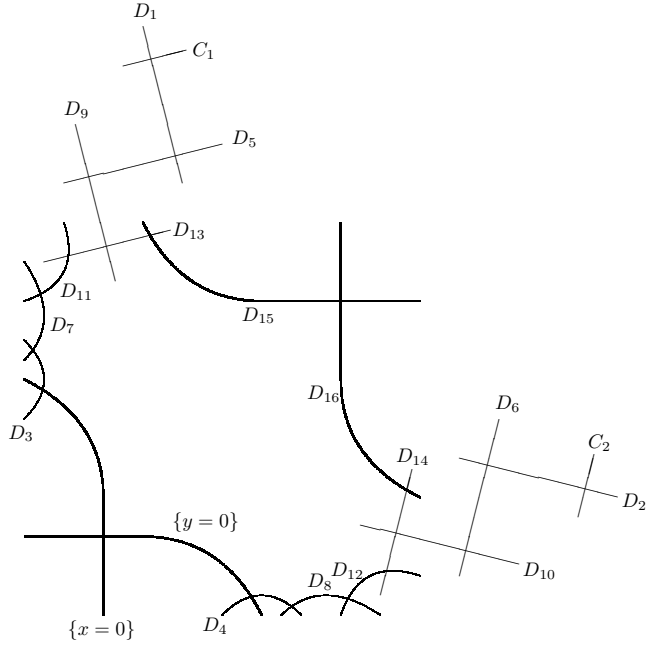
\begin{figure}[h!]
\begin{center}
\resizebox{8.5cm}{!}{
\begin{picture}(320, 340)
	
	{\thicklines
	\put(0, 60){\line(1, 0){60}}
	\put(160, 140){\line(0, 1){80}}
	\qbezier(160, 140)(160, 100)(200, 80)
	\qbezier(60, 60)(100, 60)(120, 20)
	}
	\qbezier(100, 20)(120, 40)(140, 20)
	\qbezier(130, 20)(150, 40)(180, 20)
	\qbezier(160, 20)(170, 50)(200, 40)
	\put(180, 30){\line(1, 4){16}}
	\put(170, 66){\line(4, -1){80}}
	\put(220, 40){\line(1, 4){20}}
	\put(220, 100){\line(4, -1){80}}
	\put(280, 70){\line(1, 4){8}}

	{\thicklines
	\put(40, 20){\line(0, 1){60}}
	\put(120, 180){\line(1, 0){80}}
	\qbezier(40, 80)(40, 120)(0, 140)
	\qbezier(120, 180)(80, 180)(60, 220)
	}
	\qbezier(0, 120)(20, 140)(0, 160)
	\qbezier(0, 150)(20, 170)(0, 200)
	\qbezier(0, 180)(30, 190)(20, 220)
	\put(10, 200){\line(4, 1){64}}
	\put(46, 190){\line(-1, 4){20}}
	\put(20, 240){\line(4, 1){80}}
	\put(80, 240){\line(-1, 4){20}}
	\put(50, 300){\line(4, 1){32}}

	\put(55, 325){$D_1$}
	\put(302, 75){$D_2$}
	\put(-8, 110){$D_3$}
	\put(90, 12){$D_4$}
	\put(105, 260){$D_5$}
	\put(238, 125){$D_6$}
	\put(13, 165){$D_7$}
	\put(143, 34){$D_8$}
	
	\put(20, 275){$D_9$}
	\put(252, 40){$D_{10}$}
	\put(18, 182){$D_{11}$}
	\put(155, 40){$D_{12}$}
	
	\put(75, 213){$D_{13}$}
	\put(188, 98){$D_{14}$}
	
	\put(110, 170){$D_{15}$}
	\put(143, 130){$D_{16}$}
	
	\put(85, 305){$C_1$}
	\put(285, 105){$C_2$}
	
	\put(22, 10){$\{ x = 0 \}$}
	\put(75, 65){$\{ y = 0 \}$}
	
\end{picture}
}
\end{center}\vskip-.0cm
\caption{Representation of the exceptional curves in the surface on which $\phi_n$ \eqref{phimap} acts as an automorphism. The labels $\{x=0\}$ and  $\{y=0\}$ refer to the strict transforms of the corresponding curves in  $\mathbb{P}^1 \times \mathbb{P}^1$.}
\label{fig5}\vskip-.0cm
\end{figure}

We have thus recovered the two conditions obtained from the singularity confinement analysis of the mapping \eqref{eqv} in section \ref{sec2}, and we can conclude that if the parameters $a_n$ and $b_n$ satisfy the constraints (\ref{phicond1}, \ref{phicond2}), the map $\phi_n$ can be fully regularised through (two times eight) 16 blow-ups. In other words, under the above conditions, $\phi_n$ will act as an automorphism on the (family of) surfaces obtained by glueing together all the coordinate charts used in the blow-ups. Figure \ref{fig5} depicts the exceptional curves obtained from the blow-ups, that are contained in such a surface.

Note that, as in the first example, here as well we see that the presence or absence of the term $b_n/x_n$ in the mapping does not alter the number of blow-ups (or their structure) needed for the regularisation of the mapping, which was the basic ansatz in the full-deautonomisation procedure.

Since 16 blow-ups of $\mathbb{P}^1 \times \mathbb{P}^1$ were needed to construct the surface shown in figure \ref{fig5}, its Picard group will be of rank 18 and we can take the curves $D_1, D_2, \ldots, D_{16}, C_1$ and $C_2$ as a basis. Moreover, the analysis of the correspondences between the base-points for the different blow-ups, given above, also establishes the automorphism $\phi_*$ induced by the map $\phi_n$ on the Picard group:
\begin{gather}
D_1 \to D_2 \to D_3 \to D_4 \to D_1\,,\qquad D_5 \to D_6 \to D_7 \to D_8 \to D_5\,,\label{phistD1}\\
D_9 \to D_{10} \to D_{11} \to D_{12} \to D_9\,,\qquad D_{13} \to D_{14} \to D_{13}\,,\qquad D_{15} \to D_{16} \to D_{15}\,,\label{phistD2}
\end{gather}
and
\begin{equation}
\{ y = 0 \} \to C_1 \to C_2 \to \{ x = 0 \}.
\end{equation}
Note that the $y$ dependence in this last sequence of curves corresponds exactly to the singularity pattern $\{ 0, \infty^4, 0\}$ for the mapping \eqref{eqv}.

As $\{ x = 0 \}$ is linearly equivalent to
\begin{multline}
	- D_1 + 4 D_2 - D_3 + D_4 - D_5 +4 D_6 - D_7 + 2 D_8 - D_9 +4 D_{10} - D_{11} + 3 D_{12}\\ - D_{13} + 4 D_{14} + D_{16} - C_1 + 4 C_2\,,
\end{multline}
we find that $\phi_{*}$ can be represented by a matrix of the form
$\left( \begin{array}{c|c}
	\sigma & * \\ \hline
	0 & A \\
\end{array} \right)$, where $\sigma$ is a unitary sub-matrix (as it represents the permutations (\ref{phistD1},\ref{phistD2}) of the $D$ curves) and where 
$A = \begin{pmatrix}
	0	&	-1 \\
	1	&	4
\end{pmatrix}$. Hence, it is the characteristic polynomial for $A$,
\begin{equation}
	\det(\lambda I - A) = \lambda^2 - 4 \lambda + 1,
\end{equation}
that determines the largest eigenvalue of $\phi_*$: $\lambda_*=2+\sqrt{3}$. This proves that the algebraic entropy of $\phi_n$ -- and therefore also that of \eqref{eqv} -- is positive and that this mapping is indeed non-integrable, as was surmised in section \ref{sec2}. The important thing to note however is that, here again, the characteristic polynomial for the sub-matrix $A$ coincides with that of the (only) non-trivial constraint, \eqref{phicond2}, on the parameters in the mapping. Because this is exactly the factor in the full characteristic polynomial of $\phi_*$ that decides on the non-integrability of the mapping, it becomes clear that although the absence or presence of the $b_n/x_n$ term in the mapping does not alter the surface obtained from the blow-ups -- or the action of the mapping on the Picard group for that matter -- without this term, this correspondence between the constraints on the parameters and the action on the Picard group would be broken and  it would be impossible to read-off the value of the largest eigenvalue of $\phi_*$ just from the constraints on the coefficients. This fact justifies the full-deautonomisation approach, in which one supplements the mapping with terms that do not alter the singularity pattern, but that are general enough to capture, so to speak, the full dynamics of the mapping induced on the Picard group, at the level of the parameters.

\section{Applications of the full-deautonomisation procedure}
In the preceding sections we presented the basic workings of the full-deautonomisation method and its justification based on algebro-geometric arguments. Here, in order to underline the new power the singularity confinement method gains from this approach, we present selected examples where this novel method makes it possible to detect non-integrability despite the fact that they all have confined singularities. 

In \cite{viallet}, Viallet offered an example of a mapping that possesses confined singularities but which has non-zero algebraic entropy and is thus not expected to be integrable:
\begin{equation}
x_{n+1}x_{n-1}=x_n+{1\over x_n}+a\label{fqi}
\end{equation}
(The above mapping, with $a=0$, was first published by Hone  \cite{hone} who, however, attributes its paternity to Viallet). The singularity pattern of \eqref{eqi} is $\{0, \infty, \infty^2, \infty, 0\}$. The standard deautonomisation of \eqref{fqi} consists in just assuming that $a_n$ is a function of the independent variable. Implementing the singularity analysis  in this case, while requiring that the pattern be the same as for the autonomous one (and in particular when starting from a finite $x_n$ and a vanishing $x_{n+1}$ requiring that $x_{n+7}$ be finite), we obtain the constraint $a_{n+6}=a_n$, i.e. $a_n$ should be a periodic function with period 6. 

The full-deautonomisation of \eqref{fqi} would require that we introduce functions multiplying the terms proportional to $x$ and $1/x$ in the right-hand side as well as keeping $a_n$. However it turns out that $a$ does not play any role in fixing the value of the algebraic entropy, and since $a_n=0$ is a possible solution to the constraint that was obtained in the previous paragraph, in what follows, we choose its value to be 0. We shall thus work with
\begin{equation}
x_{n+1}x_{n-1}=b_nx_n+{c_n\over x_n}.\label{fqii}
\end{equation}
We perform the singularity analysis requiring that the singularity pattern be the same as for the autonomous case and obtain the confinement constraint
\begin{equation}
b_{n+6}b_{n}c_{n+5}^2c_{n+1}^2=b_{n+4}^2b_{n+3}^4b_{n+2}^2c_{n+6}c_{n}.\label{fqiii}
\end{equation}
Inspecting equation \eqref{fqii} it is clear that there exists a global gauge freedom which allows to put either $b_n$ or $c_n$ equal to 1. Thus the confinement constraint can be expressed as either
\begin{equation}
c_{n+5}^2c_{n+1}^2=c_{n+6}c_{n}\label{fqiv}
\end{equation}
or 
\begin{equation}
b_{n+6}b_{n}=b_{n+4}^2b_{n+3}^4b_{n+2}^2.\label{fqivb}
\end{equation}
Seeking a solution to these constraints for $\log c_n$ or $\log b_n$ in the form $\lambda^n$ we obtain the characteristic equation $(\lambda^2\pm\lambda+1)(\lambda^4-\lambda^3-2\lambda^2-\lambda+1)=0$, where the plus and minus signs in the first factor correspond to \eqref{fqiv} and \eqref{fqivb} respectively. The largest root of this characteristic equation is $(1+\sqrt{17})/4+\sqrt{(1+\sqrt{17})/8}$ the approximate value of which is 2.081019, in perfect agreement with the value obtained by Viallet in his calculation of the degree growth of \eqref{fqi}. 

Next we turn to a generalisation of mapping \eqref{eqiv} which was analysed in section \ref{sec2}:
\begin{equation}
x_{n+1}+x_{n-1}={a_n\over x_n^k},\label{fqv}
\end{equation}
where $k$ is a positive integer greater than 2 (the cases $k=1$ and $k=2$ being integrable). The requirement that the singularity confines after three steps, introduces the constraint $a_{n+1}=(-1)^ka_{n-1}$. (For even $k$ this means that $a_n$ must be a periodic function with period 2 while in the case of odd $k$ the function $a_n$ must be of period 4 but involving only powers of $i$ and $-i$). In order to show the non-integrability of \eqref{fqv}, we perform a full-deautonomisation by adding all terms of the form $x_n^{-\ell}$ with $\ell=1,\cdots,k-1$ with free coefficients. However, with hindsight, as in the case analysed in section \ref{sec2}, the term $x_n^{-1}$ alone already suffices to obtain the answer we seek. We thus work with
\begin{equation}
x_{n+1}+x_{n-1}={b_n\over x_n}+{a_n\over x_n^k},\label{fqvi}
\end{equation}
and we require that the singularity pattern be $\{0, \infty^k, 0\}$, as it was in the absence of the $b_n/x_n$ term. Using the proper choice for $a_n$ we readily find the condition $b_{n+1}-kb_n+b_{n-1}=0$ which leads to the characteristic equation $\lambda^2-k\lambda+1=0$, the largest root of which is equal to $(k+\sqrt{k^2-4})/2$. A numerical estimate of the degree growth in the case $k=4$, obtained with the help of the Diophantine approximation introduced by Halburd \cite{halburd}, leads after 12 iterations to a value of 3.732052, very close to the expected one, $2+\sqrt{3}$.

The following variant of \eqref{fqv} is also interesting in the sense that it corresponds to a different singularity pattern. Indeed for
\begin{equation}
x_{n+1}+x_{n-1}=1+{a_n\over x_n^k},\label{fqvii}
\end{equation}
and $k>1$, we find the pattern $\{0, \infty^k, 1, \infty^k, 0\}$ provided $a_n$ satisfies the constraint $a_{n+2}+a_{n-2}=0$. Here again we proceed to the full-deautonomisation by adding a term inversely proportional to $x$ with a free coefficient:
\begin{equation}
x_{n+1}+x_{n-1}=1+{b_n\over x_n}+{a_n\over x_n^k}.\label{fqviii}
\end{equation}
After implementing the constraint for $a_n$ we find that, for the singularity to be confined, $b_n$ must satisfy: $b_{n+2}-k(b_{n+1}+b_{n-1})+b_{n-2}=0$, with characteristic equation $\lambda^4-k(\lambda^3+\lambda)+1=0$.  The largest root for the latter is $(k+\sqrt{k^2+8})/4+\sqrt{(k\sqrt{k^2+8}+k^2-4)/8}$, which is greater than 1 for $k>1$. Thus we expect \eqref{fqvii} to be non-integrable for $k>1$. (The case $k=1$ is a well-known integrable one, obtained from \eqref{fqviii} with $a_n=0$ and $b_n$ satisfying the constraint $b_{n+2}-b_{n+1}-b_{n-1}+b_{n-2}=0$) \cite{sincon}. 

The final example of this section is an extension of a mapping presented in \cite{redeem}
\begin{equation}
x_{n+1}x_{n-1}={x_n^k-1\over x_n^k+1},\label{fqix}
\end{equation}
where $k$ is an integer greater than 2. (The cases $k=1$ and $k=2$ are known to be integrable) \cite{papy}.

The cases of even and odd $k$ should be treated separately since their singularity structure is different. In the case of even $k$ the singularity patterns are very simple. We have $\{r, 0, -1/r\}$, where $r$ is a $k$th root of 1, and $\{s, \infty, 1/s\}$ where $s$ is a $k$th root of $-1$. For odd $k$ the singularity patterns are $\{r, 0, -1/r, \infty, -r, 0, 1/r\}$ and $\{s, \infty, 1/s\}$ where, again, $r$ and $s$ are $k$th roots of 1 and $-1$, respectively.
Just as in the case $k=4$ analysed in \cite{redeem}, the full-deautonomisation of \eqref{fqix} can be limited to the introduction of a single free function:
\begin{equation}
x_{n+1}x_{n-1}={x_n^k-a_n^k\over x_n^k+1}.\label{fqx}
\end{equation}
In the case of even $k$ we analyse the singularity $x_n=a_n$ and require that it confine in three steps, exactly as in the autonomous case. We obtain a confinement condition $a_{n+1}a_{n-1}=a_n^k$ which leads to the characteristic equation $\lambda^2-k\lambda+1=0$. The largest root of the latter is $(k+\sqrt{k^2-4})/2$ which for $k>2$ is always greater than 1. The same singularity for odd $k$ leads to a confinement constraint $a_{n+3}a_{n-3}=a_{n+2}^ka_{n-2}^k$. The corresponding characteristic equation is now $\lambda^6-k(\lambda^5+\lambda)+1=0$. While it does not appear possible to express the largest root of this equation in terms of radicals, it is easy to obtain the first terms of its asymptotic expansion, $k+1/k^3-1/k^5$, which offers a good precision already in the case $k=3$.  Thus for all $k$ greater than 2 the singularity confinement approach, combined with the full-deautonomisation procedure, offers a clear indication of the non-integrability of \eqref{fqix}, despite the presence of confined singularities.

\section{Early versus late confinement}\label{sec5}
In \cite{prsa} we examined a particular application of singularity confinement, known as {\sl late confinement}. This term however requires some clarification. When applying the singularity confinement criterion in conjunction with some deautonomisation approach, it is standard practice to confine at the very first confinement opportunity. The notion of ``very first" opportunity, however, becomes hazy when the initial mapping has several singularity patterns, leading to more than one possible deautonomisation. This is something that can be dealt with, though it requires some experience, especially as none of these deautonomisations are what we call ``late" ones, although some of them correspond to singularity patterns that are longer than others. What can serve here as a guide for the perplexed is that one has to require that every singularity pattern for the non-autonomous extension be identical to one of the autonomous mapping, but in the non-autonomous case this is not the only possibility.
As was first pointed out by Hietarinta and Viallet \cite{late}, it is possible to bypass the first confinement opportunity and still impose confinement at some later stage. However the consequence of such a ``late'' confinement is that it leads to a non-integrable system while a normal confinement would have led to an integrable one. The algebro-geometric analysis of this phenomenon was presented in \cite{prsa}.

Another possibility is that a confinement possibility appears {\sl earlier} in the singularity pattern. The existence of such a possibility is something that can be assessed already in the autonomous case. While the late confinement situation is always possible (something that is obvious once one follows the algebro-geometric analysis), an ``early'' confinement possibility is a particular situation which may arise or not, depending on the mapping. Still, several examples do exist and in what follows we shall illustrate the consequences of early and late confinement on an otherwise integrable mapping.

We start with the mapping
\begin{equation}
x_{n+1}x_{n-1}=a_n(1-x_n).\label{aqi}
\end{equation}
In the autonomous case, i.e. when $a_n=a$ for all $n$, we have the singularity pattern 
\begin{equation}
\{1, 0, a, \infty, \infty, a, 0, 1\},
\end{equation}
meaning that if we start from a finite $x_n$ and $x_{n+1}=1$, we find that $x_{n+9}$ is finite and in fact equal to $x_n$. The standard deautonomisation of \eqref{aqi} was presented in \cite{papy}. The singularity pattern, with the same structure as for the autonomous case, is now
\begin{equation}
\{1, 0, a_{n+2}, \infty, \infty, a_{n+5}a_{n+4}/a_{n+2}, 0, a_{n+7}a_{n+2}/(a_{n+5}a_{n+4})\}.
\end{equation}
The confinement constraint is $a_{n+7}a_{n+2}=a_{n+5}a_{n+4}$ and can be integrated to $\log a_n=\alpha n+\beta+\gamma(-1)^n+\delta j^n+\zeta j^{2n}$, where $j=e^{2i\pi\over3}$. Under this deautonomisation equation, \eqref{aqi} is a discrete Painlev\'e equation associated with the affine Weyl group A$_4^{(1)}$. 

However, there exists another possibility. Instead of confining after 8 steps we can choose to confine just after the third step, by taking $a_n=1$. This confinement is indeed possible but its consequence on the mapping is drastic: the mapping is not only autonomous but, starting from $x_n$ and $x_{n+1}$ we find the succession of values  $(1-x_{n+1})/x_n, (x_n+x_{n+1}-1)/(x_nx_{n+1}), (1-x_n)/x_{n+1},x_n, x_{n+1},\cdots$, i.e. the mapping is now periodic with period 5. In fact, in all studied cases we have observed the same phenomenon: if early confinement is enforced, the mapping becomes periodic (unless it becomes entirely trivial).

We turn now to the case of late confinement by deciding not to confine after 8 steps. In this case $x_{n+9}$ is infinite and the next confinement opportunity appears at the level of $x_{n+14}$. The confinement constraint is now $a_{n+10}a_{n+5}a_{n}=a_{n+8}a_{n+7}a_{n+3}a_{n+2}$, the characteristic equation for which is $\lambda^{10}-\lambda^8-\lambda^7+\lambda^5-\lambda^3-\lambda^2+1=0$, with largest root $\lambda=1.293486$. A numerical evaluation of this root using Halburd's Diophantine method yields, after 50 iterations, a value of 1.31 and the convergence to the value obtained from the characteristic polynomial is extremely slow. In the next section we shall prove however that the above largest root does indeed lead to the correct degree growth for this mapping.

One further remark should be made at this point. The late confinement we introduced at the level of $x_{n+14}$ is not the only possible one. In fact, infinitely many confinement opportunities exist  every 5 steps, i.e. for all indices of the form $n+9+5m$. A study of the algebraic entropy for successive late confinements shows that it grows monotonically and approaches an upper limit, which can be computed exactly. Starting from the characteristic polynomial for the first, standard, confinement, $P_0(\lambda)=\lambda^5-\lambda^3-\lambda^2+1$, it is easy to express the successive $P_m$ in terms of $P_0$ as
\begin{equation}
P_m(\lambda)={\lambda^{5m+5}-1\over \lambda^5-1}\Big(P_0(\lambda)-1\Big)+1.\label{aqii}
\end{equation}
Since the largest root of $P_m$ has modulus greater than 1, we find, at the limit $m\to\infty$, that this root must coincide with that of $P_0(\lambda)-1$, which is given by the equation $\lambda^3-\lambda-1=0$. We obtain thus $\lambda_{\infty}=\left((\sqrt{27/4}-\sqrt{23/4})^{1/3}+(\sqrt{27/4}-\sqrt{23/4})^{-1/3}\right)/\sqrt{3}$ (with a numerical value approximately equal to 1.324718), the logarithm of which constitutes the upper limit of the algebraic entropy of the system, one that would necessitate an infinitely delayed confinement.

\section{On the  pitfalls of gauge freedom}\label{sec6}
The analysis in the previous section was based on the special form \eqref{aqi} of the mapping, which corresponds to a precise choice of gauge. However, another gauge could for example give $x_{n+1}x_{n-1}=1- c_n x_n$, and in general, for a given mapping, it is not always clear what the optimal choice for the gauge might be. Moreover, in what follows we shall explain that a `bad' choice of gauge, combined with an inadvertent analysis of the constraints obtained from singularity confinement, may actually lead to entirely wrong conclusions. As we shall see however, this problem is by no means insurmountable.

In order to illustrate this subtle point, we start from the mapping \eqref{aqi} and use the full gauge freedom, which allows us to write it as
\begin{equation}
x_{n+1}x_{n-1}=a_n-b_nx_n.\label{aqiii}
\end{equation}
Next we repeat the singularity analysis starting from a finite $x_n$ and $x_{n+1}=a_{n+1}/b_{n+1}$, while requiring confinement after 8 steps. We readily obtain the condition
\begin{equation}
a_{n+8}a_{n+1}b_{n+5}b_{n+4}=a_{n+7}a_{n+2}b_{n+8}b_{n+1}.
\end{equation}
The case analysed in the previous section corresponds  to $b_n\equiv a_n$, but any relation between $a_n$ and $b_n$ is a priori conceivable. We can take for instance $b_n=a_n^k$, for some power $k\neq 1$, in which case the characteristic polynomial obtained from the confinement condition becomes $(k-1)(\lambda^7+1)+\lambda^6+\lambda-k(\lambda^4+\lambda^3)$, which factorises into $(\lambda-1)^2(\lambda+1)(\lambda^2+\lambda+1)\left[(k-1)(\lambda^2+1)+\lambda\right]$. While the first three factors correspond to the integrable $n$-dependence obtained in the previous section, the last factor creates a major problem. In fact, choosing $k$ appropriately, we can make the largest root of this characteristic polynomial equal to any positive number we wish. In which case, according to the conjecture presented in \cite{redeem} and which is supported by all the results in the preceding sections, this would be an indication of non-integrability for \eqref{aqiii}. This conclusion, if correct, would of course toll the bell for singularity confinement as a discrete integrability detector, since it is in direct contradiction with the basic intuition that says that the mapping \eqref{aqi} should be integrable, regardless of the gauge one chooses. 

However, this hasty conclusion is totally unwarranted. Using the transformation $x_n\to\gamma_n x_n$, the mapping \eqref{aqiii} with $b_n=a_n^k$ can be cast into the form \eqref{aqi} by choosing the gauge function $\gamma_n$ such that it satisfies
$\gamma_n = (\gamma_{n+1}\gamma_{n-1})^{1-k}$, which of course corresponds to the characteristic equation $(k-1)(\lambda^2+1)+\lambda=0$, for $\log\gamma_n$. Hence it becomes clear that the troublesome factor obtained in the above analysis is entirely due to the specific choice of gauge in the mapping \eqref{aqiii},  and that its contribution to the coefficients can be easily undone by a simple transformation of the dependent variables. Most importantly, this factor in the characteristic polynomial does not influence the degree growth of the mapping at all, as we shall explain shortly by means of an algebro-geometric analysis of \eqref{aqiii}. Moreover we shall show that, strictly speaking, such an analysis is not even necessary to solve this gauge conundrum.

In fact, as will become clear from the algebro-geometric analysis of \eqref{aqiii}, the key to successfully detecting contributions to the confinement constraints that are due solely to gauge freedom, lies in the late confinements of the mapping. Let us for example proceed to the late confinement for \eqref{aqiii} obtained by requiring that the singularity confine at the level of $x_{n+14}$. We obtain the constraint $a_{n+13}a_{n+1}b_{n+10}b_{n+9}b_{n+5}b_{n+4}=a_{n+12}a_{n+7}a_{n+2}b_{n+13}b_{n+1}$ which, when $b_n=a_n$, reduces to the one we obtained in the previous section. Taking again  $b_n=a_n^k$ we find the characteristic polynomial $(k-1)(\lambda^{12}+1)+\lambda^{11}+\lambda^6+\lambda-k(\lambda^9+\lambda^8+\lambda^4+\lambda^3)$ which factorises into $(\lambda^{10}-\lambda^8-\lambda^7+\lambda^5-\lambda^3-\lambda^2+1)$ and $\left[(k-1)(\lambda^2+1)+\lambda\right]$. The first of these factors is the one obtained when we implemented late confinement on \eqref{aqi}. More importantly however, the second factor is common to both the standard and the late confinement for \eqref{aqiii} and its role is now clear: it is induced by the special choice of gauge, $b_n=a_n^k$. 
Such a gauge-induced factor has no bearing whatsoever on the integrability of the mapping and should be disregarded when analysing the characteristic equation. 

Thus, the correct prescription for the implementation as an integrability detector of singularity confinement, with full-deautonomisation, can be codified as follows:  one should not only perform the analysis for the standard confinement but also for the late ones (although, often, the first late confinement will suffice). Any common factor in the characteristic polynomials of the standard and late confinements indicates the existence of a gauge freedom. As such, it does not play any role in the integrability of the mapping and should be disregarded. 

\begin{figure}[h!]
\begin{center}
\resizebox{8.5cm}{!}{
\begin{picture}(300, 280)
	{\thicklines
	\put(0, 30){\line(1, 0){100}}
	\put(0, 90){\line(1, 0){100}}
	\put(20, 10){\line(0, 1){100}}
	\put(80, 10){\line(0, 1){100}}
	}
	
	\put(80, 90){\circle*{5}}
	\put(20, 50){\circle*{5}}

	\put(5, 0){$x_n = 0$}
	\put(65, 0){$s_n = 0$}
	\put(105, 28){$y_n = 0$}
	\put(105, 88){$t_n = 0$}

	\put(50, 130){$\downarrow$}

	{\thicklines
	\put(0, 180){\line(1, 0){100}}
	\put(0, 240){\line(1, 0){50}}
	\put(20, 160){\line(0, 1){100}}
	\put(80, 160){\line(0, 1){50}}
	\qbezier(80, 210)(80, 225)(100, 240)
	\qbezier(50, 240)(65, 240)(80, 260)
	}
	\put(60, 260){\line(1, -1){40}}
	\put(10, 200){\line(1, 0){20}}

	\put(40, 180){\circle*{5}}
	\put(35, 168){$P_n$}

	{\thicklines
	\put(200, 30){\line(1, 0){100}}
	\put(200, 90){\line(1, 0){100}}
	\put(220, 10){\line(0, 1){100}}
	\put(280, 10){\line(0, 1){100}}
	}
	
	\put(280, 90){\circle*{5}}
	\put(240, 30){\circle*{5}}

	\put(250, 130){$\downarrow$}

	{\thicklines
	\put(200, 180){\line(1, 0){100}}
	\put(200, 240){\line(1, 0){50}}
	\put(220, 160){\line(0, 1){100}}
	\put(280, 160){\line(0, 1){50}}
	\qbezier(280, 210)(280, 225)(300, 240)
	\qbezier(250, 240)(265, 240)(280, 260)
	}
	\put(260, 260){\line(1, -1){40}}
	\put(240, 170){\line(0, 1){20}}

	\put(220, 200){\circle*{5}}
	\put(223, 202){$Q_{n+1}$}

	\put(150, 50){$\dashrightarrow$}
	\put(145, 200){$\xrightarrow[\psi_n]{\sim}$}

\end{picture}
}
\end{center}\vskip-.0cm
\caption{Diagram showing the base-points in the two blow-ups for the map $\psi_n$ (on the left) and those for the two blow-ups for the inverse map $\psi^{-1}_n$ (on the right), together with the resulting exceptional curves.}
\label{fig6}\vskip-.0cm
\end{figure}
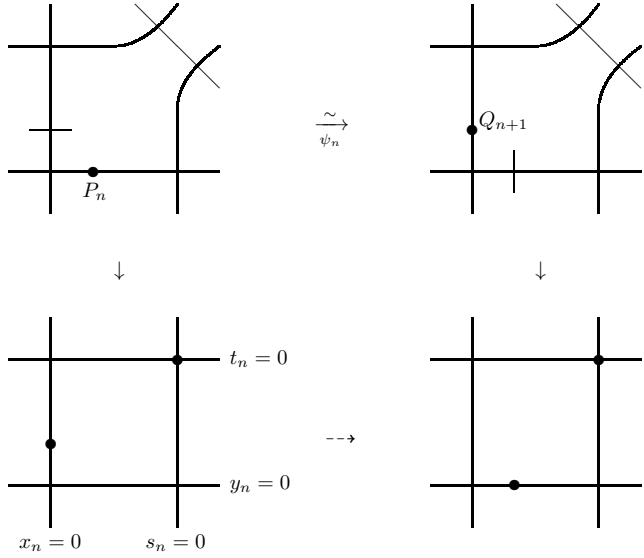

Lest this argument appear arbitrary we proceed now to offer its algebro-geometric justification.
As before, we interpret the mapping \eqref{aqiii} as a birational map on $\mathbb{P}^1 \times \mathbb{P}^1 = (x_n, y_n) \cup (x_n, t_n) \cup (s_n, y_n) \cup (s_n, t_n)$ (with $s_n = 1 / x_n, t_n = 1 / y_n$) :
\begin{equation}
	\psi_n \colon\quad \mathbb{P}^1 \times \mathbb{P}^1 \dashrightarrow \mathbb{P}^1 \times \mathbb{P}^1,
	\qquad (x_n, y_n) \mapsto (x_{n+1}, y_{n+1}) = \left( y_n, \frac{a_n - b_n y_n}{x_n} \right).\label{psimap}
\end {equation}
This map has two indeterminate points, $(s_n, t_n) = (0, 0)$ and $(x_n, y_n) = \left( 0, \frac{a_n}{b_n} \right)$, which can both be regularised with a single blow-up. Indeed, the blow-up at the first point
\begin{equation}
(s_n, t_n) \leftarrow \left( s_n, \frac{t_n}{s_n} \right) \cup \left( \frac{s_n}{t_n}, t_n \right),
\end{equation}
immediately yields the mapping $y_{n+1} = \frac{s_n}{t_n} (a_n t_n - b_n)$, and the blow-up at the second point
\begin{equation}
\left( x_n, y_n - \frac{a_n}{b_n} \right) \leftarrow \left( x_n, \frac{y_n - \frac{a_n}{b_n}}{x_n} \right) \cup \left( \frac{x_n}{y_n - \frac{a_n}{b_n}}, y_n - \frac{a_n}{b_n} \right),
\end{equation}
the mapping $y_{n+1} = - b_n \frac{y_n - \frac{a_n}{b_n}}{x_n}$, both of them well-defined.
The inverse map
\begin{equation}
	\psi^{-1}_n \colon\quad \mathbb{P}^1 \times \mathbb{P}^1 \dashrightarrow \mathbb{P}^1 \times \mathbb{P}^1,
	\qquad (x_{n+1}, y_{n+1}) \mapsto (x_n, y_n) = \left( \frac{a_n - b_n x_n}{y_n}, x_{n+1} \right),
\end{equation}
also has two indeterminate points, $(s_{n+1}, t_{n+1}) = (0, 0)$ and $(x_{n+1}, y_{n+1}) = \left( \frac{a_n}{b_n}, 0 \right)$, which can also be regularised with just one blow-up each. Their respective coordinate charts are:
\begin{equation}
(s_{n+1}, t_{n+1}) \leftarrow \left( s_{n+1}, \frac{t_{n+1}}{s_{n+1}} \right) \cup \left( \frac{s_{n+1}}{t_{n+1}}, t_{n+1} \right),
\end{equation}
and
\begin{equation}
\left( x_{n+1} - \frac{a_n}{b_n}, y_{n+1} \right) \leftarrow \left( x_{n+1} - \frac{a_n}{b_n}, \frac{y_{n+1}}{x_{n+1} - \frac{a_n}{b_n}} \right) \cup \left( \frac{x_{n+1} - \frac{a_n}{b_n}}{y_{n+1}}, y_{n+1} \right).
\end{equation}
These blow-ups are depicted in figure \ref{fig6}, together with the resulting exceptional curves. Note that $\psi_{n-1}^{-1}$ and $\psi_{n}$ share the indeterminate point $(s_{n}, t_{n}) = (0, 0)$, for all $n$, which can therefore be resolved by one and the same blow-up (to be performed at each value of $n$).

If we denote the second indeterminate point of $\psi_{n-1}^{-1}$ as
\begin{equation}
P_n~:\quad(x_n, y_n) = \left( \frac{a_{n-1}}{b_{n-1}}, 0 \right),
\end{equation}
and that of $\psi_{n+1}$ as
\begin{equation}
Q_{n+1}~:\quad(x_{n+1}, y_{n+1}) = \left( 0, \frac{a_{n+1}}{b_{n+1}} \right),
\end{equation}
then the possible scenarios for regularising the map $\psi_n$, for all $n$, can be summarised as:
\begin{equation}
	 \psi_{n+5\ell} \cdots \psi_n(P_n) = Q_{n+1+5\ell},\label{psiclosure}
\end{equation}
for some non-negative integer $\ell$. 

In particular, the case $\ell=0$, i.e. $\psi_n(P_n)\equiv Q_{n+1}$, will be seen to correspond to the early confinement discussed in section \ref{sec5}, $\ell=1$ to the standard confinement and the cases $\ell\geq2$ to all possible late confinements for this map. In order to analyse these different cases in all generality we first define the points
\begin{gather}
T^{(1)}_n(\alpha): (s_n, y_n) = (\alpha, 0)\,,\qquad T^{(2)}_n(\beta): (x_n, y_n) = (0, \beta)\,,\qquad T^{(3)}_n(\gamma): (x_n, t_n) = (\gamma, 0)\,,\\
T^{(4)}_n(\delta): \left( s_n, - \frac{t_n}{s_n} \right) = (0, \delta)\,,\qquad T^{(5)}_n(\varepsilon): (s_n, t_n) = (0, \varepsilon)\,,
\end{gather}
which are related through the action of $\psi_n$:
\begin{gather}
\psi_{n}(T^{(1)}_n(\alpha)) = T^{(2)}_{n+1}(a_n \alpha)\,,\qquad \psi_{n}(T^{(2)}_n(\beta)) = T^{(3)}_{n+1}(\beta)\,,\qquad \psi_{n}(T^{(3)}_n(\gamma)) = T^{(4)}_{n+1} \left( \frac{\gamma}{b_n} \right)\,,\\
\psi_{n}(T^{(4)}_n(\delta)) = T^{(5)}_{n+1} \left( \frac{\delta}{b_n} \right)\,,\qquad\psi_{n}(T^{(5)}_n(\varepsilon)) = T^{(1)}_{n+1}(\varepsilon)\,.
\end{gather}
We have that $P_n = T^{(1)}_n \left( \frac{b_{n-1}}{a_{n-1}} \right)$ and $Q_{n+1} = T^{(2)}_{n+1} \left( \frac{a_{n+1}}{b_{n+1}} \right)$ and the case of early confinement ($\ell=0$) corresponds to
\begin{equation}
	T^{(1)}_n \left( \frac{b_{n-1}}{a_{n-1}} \right)
	\mapsto
	T^{(2)}_{n+1} \left( \frac{a_{n}b_{n-1}}{a_{n-1}} \right)
	= Q_{n+1}\,,\label{psiearly}
\end{equation}
which yields the condition:
\begin{equation}
	\frac{a_{n}b_{n-1}}{a_{n-1}} = \frac{a_{n+1}}{b_{n+1}}.
\end{equation}
Note that when the requirement \eqref{psiearly} is satisfied, the map $\psi_n$ becomes well-defined after a mere three blow-ups.

When $b_n=a_n$, i.e. in the case of the mapping \eqref{aqi}, this amounts to requiring $a_n=1$ for all $n$, which is indeed the early confinement condition discussed in section \ref{sec5}. On the other hand, if we impose $b_n = a_n^k$, we find the condition
\begin{equation}
	a^{k-1}_{n+1} a_n a^{k-1}_{n-1} = 1,
\end{equation}
which can be solved in terms of the characteristic polynomial
\begin{equation}
	(k - 1)(\lambda^2+1) + \lambda,
\end{equation}
which is identical to that for the gauge transformation found above.

For the case $\ell=1$, the condition that needs to be fulfilled for the map to become well-defined is

\begin{multline}
	T^{(1)}_n \left( \frac{b_{n-1}}{a_{n-1}} \right)
	\mapsto
	T^{(2)}_{n+1} \left( \frac{a_{n}b_{n-1}}{a_{n-1}} \right)
	\mapsto
	T^{(3)}_{n+2} \left( \frac{a_{n}b_{n-1}}{a_{n-1}} \right)
	\mapsto
	T^{(4)}_{n+3} \left( \frac{a_{n}b_{n-1}}{b_{n+2}a_{n-1}} \right)\\
	\mapsto
	T^{(5)}_{n+4} \left( \frac{a_{n}b_{n-1}}{b_{n+3}b_{n+2}a_{n-1}} \right)
	\mapsto
	T^{(1)}_{n+5} \left( \frac{a_{n}b_{n-1}}{b_{n+3}b_{n+2}a_{n-1}} \right)
	\mapsto
	T^{(2)}_{n+6} \left( \frac{a_{n+5}a_{n}b_{n-1}}{b_{n+3}b_{n+2}a_{n-1}} \right)
	= Q_{n+6}\,,
\end{multline}
which will only take place after 8 blow-ups since the intermediate points $T_{n+1}^{(2)}, \cdots, T_{n+5}^{(1)}$ will each require a separate blow-up as well. The condition on the parameters that arises from the above constraint is of course
\begin{equation}
	\frac{a_{n+5}a_{n}b_{n-1}}{b_{n+3}b_{n+2}a_{n-1}} = \frac{a_{n+6}}{b_{n+6}},
\end{equation}
which, in case $b = a^k$, yields
\begin{equation}
	a^{k-1}_{n+6} a_{n+5} a_n a^{k-1}_{n-1} = a^k_{n+3} a^k_{n+2}.
\end{equation}
The characteristic polynomial for this relation is
\begin{equation}
	(k - 1)\lambda^7 + \lambda^6 - k \lambda^4 - k \lambda^3 + \lambda + (k - 1) = (\lambda-1)^2(\lambda+1)(\lambda^2+\lambda+1)\left[(k-1)(\lambda^2+1)+\lambda\right],
\end{equation}
as found above for the standard confinement. The corresponding condition for the mapping \eqref{aqi} is recovered at $k=1$.

For general $\ell \ge 2$ the closure condition \eqref{psiclosure} takes the form
\begin{equation}
	T^{(1)}_n \left( \frac{b_{n-1}}{a_{n-1}} \right)
	\mapsto
	\cdots
	\mapsto
	T^{(2)}_{n+5\ell+1} \left( \frac{a_{n}b_{n-1}}{a_{n-1}} \prod^{\ell}_{j=1} \frac{a_{n+5j}}{b_{n+5j-2}b_{n+5j-3}} \right)\label{psigencondpt}
	= Q_{n+1+5\ell},
\end{equation}
which amounts to the constraint
\begin{equation}
	\frac{a_{n}b_{n-1}}{a_{n-1}} \prod^{\ell}_{j=1} \frac{a_{n+5j}}{b_{n+5j-2}b_{n+5j-3}} = \frac{a_{n+5\ell+1}}{b_{n+5\ell+1}}\label{psigencond}
\end{equation}
on the parameters. When we put $b = a^k$ this yields
\begin{equation}
	a^{k-1}_{n+5\ell+1} a_n a^{k-1}_{n-1} \prod^{\ell}_{j=1} \frac{a_{n+5j}}{a^k_{n+5j-2}a^k_{n+5j-3}} = 1\,,
\end{equation}
which can be solved in terms of the characteristic polynomial
\begin{equation}
	g_{\ell}(\lambda) := 
	(k - 1)\left(\lambda^{5\ell+2} +1\right) + \lambda + \lambda^3 (\lambda^3 - k \lambda - k) \sum^{\ell-1}_{j=0}\lambda^{5j}.
\end{equation}
At $\ell=2$ we find  $g_2(\lambda) = (k-1)(\lambda^{12}+1)+\lambda^{11}+\lambda^6+\lambda-k(\lambda^9+\lambda^8+\lambda^4+\lambda^3)$, which is nothing but the characteristic polynomial found above for the first late confinement. Note that regularisation of the map $\psi_n$ under the condition \eqref{psigencond} will be achieved after exactly $5\ell+3$ blow-ups of $\mathbb{P}^1 \times \mathbb{P}^1$. The resulting surface, of Dynkin type $A_4^{(1)}$, is depicted in figure 7.

\begin{figure}[h!]
\begin{center}
\resizebox{8cm}{!}{
\begin{picture}(300, 245)

	{\thicklines
	\put(0, 60){\line(1, 0){200}}
	\put(0, 180){\line(1, 0){100}}
	\put(40, 20){\line(0, 1){200}}
	\put(160, 20){\line(0, 1){100}}
	\qbezier(160, 120)(160, 150)(200, 180)
	\qbezier(100, 180)(130, 180)(160, 220)
	}
	\put(120, 220){\line(1, -1){80}}
	
	\put(80, 40){\line(0, 1){40}}
	\put(90, 40){\line(0, 1){40}}
	\put(100, 40){\line(0, 1){40}}
	\put(110, 40){\line(0, 1){40}}
	
	\put(75, 30){$C_1$}
	\put(105, 30){$C_{5\ell+1}$}

	\put(20, 100){\line(1, 0){40}}
	\put(20, 110){\line(1, 0){40}}
	\put(20, 120){\line(1, 0){40}}
	\put(20, 130){\line(1, 0){40}}
	
	\put(5, 130){$C_{2}$}
	\put(0, 95){$C_{5\ell+2}$}

	\put(70, 160){\line(0, 1){40}}
	\put(80, 160){\line(0, 1){40}}
	\put(90, 160){\line(0, 1){40}}
	
	\put(90, 205){$C_3$}
	\put(60, 205){$C_{5\ell-2}$}

	\put(155, 155){\line(1, 1){30}}
	\put(145, 165){\line(1, 1){30}}
	\put(135, 175){\line(1, 1){30}}
	
	\put(187, 187){$C_4$}
	\put(167, 207){$C_{5\ell-1}$}

	\put(140, 90){\line(1, 0){40}}
	\put(140, 100){\line(1, 0){40}}
	\put(140, 110){\line(1, 0){40}}
	
	\put(185, 110){$C_{5\ell}$}
	\put(185, 85){$C_5$}

	\put(205, 55){$D_1$}
	\put(25, 20){$D_2$}
	\put(0, 185){$D_3$}
	\put(110, 225){$D_4$}
	\put(165, 20){$D_5$}

\end{picture}
}
\end{center}\vskip-.0cm
\caption{Representation of the exceptional curves in the surface on which $\psi_n$ \eqref{psimap} acts as an automorphism. The labels $\{x=0\}$ and  $\{y=0\}$ refer to the strict transforms of the corresponding curves in  $\mathbb{P}^1 \times \mathbb{P}^1$.}
\label{fig7}\vskip-.0cm
\end{figure}
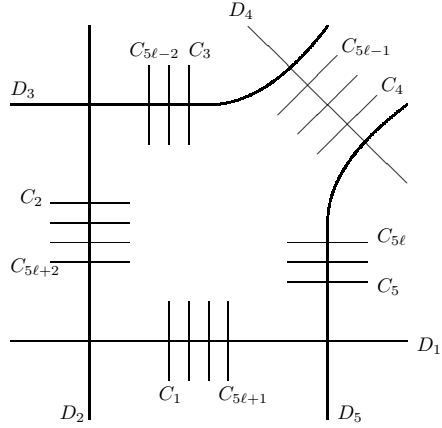

The Picard group of this surface has rank $5\ell+5$ and we choose $D_1, D_2, D_3, D_4, D_5, C_1, C_2, \ldots, C_{5\ell}$ as its basis. The map $\psi_n$ induces the following automorphism $\psi_*$ on the Picard group:
\begin{gather}
D_1 \to D_2 \to D_3 \to D_4 \to D_5 \to D_1\,,\label{psiD}\\
\left\{ y = \frac{a}{b} \right\} \to C_1 \to \cdots \to C_{5\ell} \to C_{5\ell+1} \to C_{5\ell+2} \to \left\{ x = \frac{a}{b} \right\}\,,\label{psisingpat}
\end{gather}
where the curve $C_{5\ell+1}$ is linearly equivalent to
\begin{equation}
	- D_1 + D_3 + D_4 + \sum^{\ell-1}_{j=0}(- C_{5j+1} + C_{5j+3} + C_{5j+4}).
\end{equation}
The sequence \eqref{psisingpat} corresponds to a general  singularity pattern of length $5\ell+3$, as found for the general (i.e. early, standard and late) confinements of the mappings \eqref{aqi} and \eqref{aqiii}.

The automorphism $\psi_*$ can be represented by $\left( \begin{array}{c|c}
	\sigma & * \\ \hline
	0 & A \\
\end{array} \right)$, where $\sigma$ is a $5\times5$ unitary matrix that corresponds to the permutation of the $D$ curves in \eqref{psiD} (which has period 5) and where the $5\ell\times5\ell$ matrix $A$ is given by:
\begin{equation}
A = \begin{pmatrix}
	0 &   &   &   &   &   &   &   &   & -1 \\
	1 & 0 &   &   &   &   &   &   &   & 0 \\
	0 & 1 & 0 &   &   &   &   &   &   & 1 \\
	  & 0 & 1 & 0 &   &   &   &   &   & 1 \\
	  &   & 0 & 1 &   &   &   &   &   & 0 \\
	  &   &   & 0 &   &   &   &   &   &   \\
	  &   &   &   & \ddots    &   &   &   &   & \vdots \\
	  &   &   &   &   &   &   &   &   & -1 \\
	  &   &   &   &   &   &   &   &   & 0 \\
	  &   &   &   &   &   &   & 0 &   & 1 \\
	  &   &   &   &   &   &   & 1 & 0 & 1 \\
	  &   &   &   &   &   &   & 0 & 1 & 0
\end{pmatrix}.
\end{equation}
As explained before, it is the characteristic polynomial of this matrix $A$,
\begin{equation}
	f_{\ell}(\lambda) := \det(\lambda I - A) = \lambda^{5\ell} + (- \lambda^3 - \lambda^2 + 1) \sum^{\ell-1}_{j=0} \lambda^{5j},
\end{equation}
that will determine the possible integrability or non-integrability of the map $\psi_n$. At $\ell=1$, we find $f_1(\lambda) = (\lambda+1) (\lambda^2+\lambda+1) (\lambda-1)^2 $, which is nothing but the characteristic polynomial for the integrable case of \eqref{aqi}. Discarding the trivial periodic case $\ell=0$, this is in fact the only integrable case for the map $\psi_n$ as well. It is easily checked that $f_\ell(1)=1-\ell$, which is negative for all $\ell\geq2$. Hence, for any late confinement, the automorphism $\psi_*$ will have an eigenvalue that is greater than 1 and will therefore always have positive algebraic entropy and will thus be non-integrable. Note that, for $\ell=2$, this (largest) eigenvalue is exactly the one we found for the first late confinement in section \ref{sec5}. 

It is clear that for this map $\psi_n$, the characteristic polynomial obtained from the automorphism on the Picard group does not coincide with the one obtained from the constraints on the parameters. However, one can readily check that the characteristic polynomials $g_{\ell}(\lambda)$ obtained from the constraints on the parameters $a_n$ and $b_n=a_n^k$, can be factorized in terms of $f_\ell(\lambda)$ and the polynomial $(k - 1) (\lambda^2+1) + \lambda$ which characterizes the gauge transformation between the mappings \eqref{aqiii} and \eqref{aqi} : 
\begin{equation}
	g_{\ell}(\lambda) =  f_{\ell}(\lambda)\,\left[(k - 1) (\lambda^2+1) + \lambda\right].
\end{equation}

Obviously, the automorphism $\psi_*$ on the Picard group is completely oblivious to the effect of gauge transformations: the curves that make up the Picard group are only defined up to linear equivalence and the only effect a gauge transformation will have is that of a dilation of the base points of the blow-ups that go into the construction of the surface. On the other hand, because of the latter effect, the conditions \eqref{psigencondpt} or \eqref{psigencond} will feel the influence of the gauge transformation, which explains why the polynomials $f_\ell(\lambda)$ are free of these spurious factors whereas the poynomials $g_\ell(\lambda)$ are not. This justifies the `full-deautonomisation cum late confinement' approach set forward in this paper. However, since any spurious gauge freedom is bound to show up in the form of a common factor in the characteristic polynomials for the standard and late confinements (and, if this possibility exists, in any early confinement), there is no need to perform a detailed algebro-geometric analysis such as that performed here, in order to be able to discard the influence of excessive gauge freedom when assesing the integrability or non-integrability of a mapping with the criterion we propose.

\section{Late confinement of non-integrable mappings}
In section \ref{sec5} we have considered the deautonomisation of an integrable mapping and we examined the consequences of both early and late confinements, `late' being defined as corresponding to singularity patterns that are longer than that for the autonomous case. However, the possibility for late confinement does not only exist in the integrable cases: whenever a mapping confines, late confinement opportunities always exist. In this section we shall illustrate this fact through a few examples, examining along the way the consequences of this late confinement on the algebraic entropy.

We start with the example used in section \ref{sec2} to illustrate the full-deautonomisation procedure:
\begin{equation}
x_{n+1}+x_{n-1}={b_n\over x_n}+{a_n\over x_n^4}.\label{bqi}
\end{equation}
Let us for the time being neglect the term $b/x$ and study the confinement associated with the term $1/x^4$. As we have seen, the standard singularity pattern $\{0, \infty^4, 0\}$ leads to an equation for $a_n$ of the form $a_{n+1}-a_{n-1}=0$ and a simple gauge allows us to take $a_n=1$. However, more confinement opportunities do exist: the next one corresponds to the pattern $\{0, \infty^4, 0, \infty^4, 0\}$, which results in an equation for $a_n$ of the form $a_{n+2}-a_n+a_{n-2}=0$ ; the one after that will be $\{0, \infty^4, 0, \infty^4, 0, \infty^4, 0\}$, with $a_{n+3}-a_{n+1}+a_{n-1}-a_{n-3}=0$ and so on. In order to obtain the algebraic entropy of these confining mappings we introduce the term $b/x$. As we have seen, the standard singularity pattern $\{0, \infty^4, 0\}$ -- corresponding to $a_n=1$ -- leads to an equation for $b_n$ of the form $b_{n+1}-4b_n+b_{n-1}=0$ and the characteristic polynomial $P_0^0(\lambda)= \lambda^2-4\lambda+1$. Now let us suppose that $b_n$ does not satisfy this equation and thus that the singularity is not confined after three steps. The next confinement opportunity corresponds now to the pattern $\{0, \infty^4, 0, \infty, 0, \infty^4, 0\}$ and yields an equation for $b_n$ of the form $b_{n+3}-4b_{n+2}+b_{n+1}-b_{n}+b_{n-1}-4b_{n-2}+b_{n-3}=0$.  The resulting characteristic polynomial is $P_0^1(\lambda)=\lambda^6-4\lambda^5+\lambda^4-\lambda^3+\lambda^2-4\lambda+1$ with largest root approximately equal to 3.805962. Again, more confinement opportunities do exist, leading to patterns containing an arbitrary number of repetitions of the block $\{0, \infty^4, 0, \infty\}$ ending by $\{0, \infty^4, 0\}$ and the associated equation for $b_n$. 

As in section \ref{sec5} we can explicitly obtain the limit of the algebraic entropy for successively delayed confinements. The first simple case corresponds to the confinement being delayed because the coefficient $a_n$ does not satisfy the lower constraint equations. In this case, starting with the characteristic polynomial for the regular confinement $P_0^0(\lambda)= \lambda^2-4\lambda+1$, we find for the $m$-times delayed one, the expression
\begin{equation}
P_m^0(\lambda)={\lambda^{2m+2}-1\over \lambda^2-1}\Big(P_0^0(\lambda)-1\Big)+1.\label{bqii}
\end{equation}
At the limit $m\to\infty$ the largest root of $P_m^0(\lambda)$ is given by that of $P_0(\lambda)-1$, which is exactly 4. 

The second simple case corresponds to $a_n=1$, the confinement this time being delayed because $b_n$ does not satisfy the lower constraint equations.  Here we find for the $k$-times delayed characteristic polynomial the expression
\begin{equation}
P_0^k(\lambda)={\lambda^{4k}-1\over \lambda^4-1}\Big(P_0^1(\lambda)-P_0^0(\lambda)\Big)+P_0^0(\lambda),\label{bqiii}
\end{equation}
or an equivalent one noticing that $P_0^1(\lambda)-P_0^0(\lambda)=\lambda^3(\lambda P_0^0(\lambda)-1)$. Thus at the limit $k\to\infty$ the largest root of $P_0^k(\lambda)$ is given by that of $\lambda P_0(\lambda)-1$, approximately equal to 3.8063007. However, it is possible to have a situation where the confinement is delayed $m$ times because of the choice of $a_n$ and a further $k$ times because of $b_n$. The corresponding singularity pattern has the following structure: the basic block $\{0, \infty^4\}$ is repeated $m+1$ times ending by a 0, forming thus what we shall call the block-$m$ ; the block-$m$ at the end of which a simple $\infty$ is added is then repeated $k$ times ending by a last block-$m$. The corresponding characteristic polynomial becomes thus
\begin{equation}
P_m^k(\lambda)={\lambda^{(2m+4)k}-1\over \lambda^{2m+4}-1}\Big(\lambda^{2m+3}(\lambda P_m^0(\lambda)-1)\Big)+P_m^0(\lambda).\label{bqiv}
\end{equation}

The second example is related to the H-V mapping, already presented in the introduction and which was treated within the full-deautonomisation approach in \cite{redeem}. In order to illustrate the effect of a late confinement we shall work with the form
\begin{equation}
x_{n+1}+x_{n-1}=x_n+{b_n\over x_n}+{a_n\over x_n^2}.\label{cqi}
\end{equation}
The standard singularity pattern of  \eqref{cqi} is $\{0, \infty^2, \infty^2, 0\}$ where the meaning of powers in the infinities was explained in section \ref{sec2}. The confinement condition on $a_n$ is just $a_{n+3}=a_n$, i.e. $a_n$ is periodic with period 3, while for $b_n$ we obtain the equation $b_{n+3}-2b_{n+2}-2b_{n+1}+b_n=0$. The characteristic polynomial of the latter is $P_0^0(\lambda)=\lambda^3-2\lambda^2-2\lambda+1$, the largest root of which is $(3+\sqrt5)/2$ (approximately equal to 2.618034),  its logarithm giving the algebraic entropy of the mapping. A first late confinement opportunity, when $a_n$ does not satisfy the first confinement condition, is related to the singularity pattern $\{0, \infty^2, \infty^2, 0, \infty^2, \infty^2, 0\}$ and leads to two constraints. The one on $a_n$ is $a_{n+6}-a_{n+3}+a_n=0$ which means that $a_n$ can be expressed as a sum of powers of the ninth roots of unity $e^{\pm{i\pi\over9}}$, $e^{\pm{2i\pi\over9}}$ and $e^{\pm{4i\pi\over9}}$. (Notice that the choice $a_n=1$ is not possible at this stage, contrary to the case of the standard H-V mapping). For $b_n$ we obtain the equation $b_{n+6}-2b_{n+5}-2b_{n+4}+b_{n+3}-2b_{n+2}-2b_{n+1}+b_n=0$ and a characteristic polynomial $P_1^0(\lambda)=\lambda^6-2\lambda^5-2\lambda^4+\lambda^3-2\lambda^2-2\lambda+1$. Its largest root  is approximately equal to 2.727069. A numerical estimate of the degree growth using Halburd's Diophantine approximation method gives, after 12 iterations, a value of 2.727167 converging nicely towards the value of the largest root. 

As in the previous cases the upper bound for the algebraic entropy for an infinitely delayed  confinement due to $a_n$ can be easily obtained. Starting from the characteristic polynomial $P_0^0(\lambda)=\lambda^3-2\lambda^2-2\lambda+1$, for the shortest confinement we obtain for the $m$-times delayed one the expression
\begin{equation}
P_m^0(\lambda)={\lambda^{3m+3}-1\over \lambda^3-1}\Big(P_0^0(\lambda)-1\Big)+1,\label{cqii}
\end{equation}
and using the same reasoning as in the previous cases we find that the largest root of $P_m(\lambda)$ for $m\to\infty$ is given by the largest root of $P_0(\lambda)-1$ which is equal to $1+\sqrt3$. The logarithm of this number constitutes the upper bound for the algebraic entropy for the H-V mapping in case of a late confinement. 
However, it is always possible to satisfy the constraint for $a_n$, for instance taking $a_n=1$, and delay confinement $k$ times  by not satisfying the first constraints on $b_n$. In this case the singularity pattern comprises a block  $\{0, \infty^2, \infty^2, 0, \infty, \infty\}$ repeated $k$ times ending by a block $\{0, \infty^2, \infty^2, 0\}$. The characteristic polynomial is given by
\begin{equation}
P_0^k(\lambda)={\lambda^{6k}-1\over \lambda^6-1}\Big(\lambda^2P_0^0(\lambda)-\lambda-1\Big)\lambda^4+P_0^0(\lambda).\label{cqiii}
\end{equation}
For $k=1$ we find that the largest root of $P_0^1$ is approximately equal to 2.67856. When $k\to\infty$, the largest root is that of the polynomial $\lambda^2P_0^0(\lambda)-\lambda-1$ which turns out to be approximately equal to 2.678712. 

Our final example is based on the extension of the H-V mapping, introduced in \cite{kanki} and analysed also in \cite{redeem},
\begin{equation}
x_{n+1}+x_{n-1}=x_n+{b_n\over x_n}+{a_n\over x_n^k},\label{cqiv}
\end{equation}
where we assume that the exponent $k$ is an odd number.
As was shown in \cite{redeem} a sufficient condition for \eqref{cqiii} to possess the same confined singularity pattern as the standard H-V mapping is that $a_n=(-1)^n$. When this condition is implemented we find that $b_n$ must satisfy the confinement relation $b_{n+3}-k(b_{n+2}+b_{n+1})+b_n=0$ leading to the characteristic polynomial $\lambda^3-k(\lambda^2+\lambda)+1$.
On the other hand, taking $a_n=1$ leads to a mapping with unconfined singularities. In \cite{kanki} Kanki and collaborators have derived the exact value of the largest root of the characteristic polynomial for this unconfined case, obtaining the expression $(k+\sqrt{k(k+4)})/2$. It is interesting that this value can be easily obtained within the late confinement approach. In fact, an infinitely delayed confinement is nothing but a case with unconfined singularities. Following the derivations presented above it is easy to convince oneself that the largest root of the characteristic polynomial for infinitely delayed confinement is given by the largest root of the polynomial corresponding to the shortest confinement minus 1. In our case this is the largest root of the equation $\lambda^2-k(\lambda+1)=0$ which leads, as expected, to the very same value obtained by Kanki and collaborators. Thus it is possible, using singularity confinement, to obtain the algebraic entropy of a mapping which does not confine, provided it possesses one special confining case.

\section{Summary and Conclusions}
Singularity confinement is the discrete analogue of the Painlev\'e property. Just as the latter ensures that the solution of a differential equation can define a function by eschewing multivaluedness-inducing singularities, the former ensures that a mapping is well-defined by resolving all indeterminacies. However, discrete integrability is in some sense more demanding that its continuous counterpart since it is intimately related to the growth properties of the solution of a given system. Singularity confinement alone might, in principle, be thought to be able to guarantee the slow growth that accompanies the integrable character of a rational mapping, since it is based on the simplification of the relevant factors that appear during the iteration of the mapping. This, together with the observation that all known mappings integrable through spectral methods possess confined singularities, was what led two of the present authors, in collaboration with V. Papageorgiou, to advance the hypothesis that the singularity confinement property could be used as an integrability detector \cite{sincon}. 

However, the usefulness of singularity confinement as a discrete integrability criterion was questioned by the discovery of a counterexample -- the (in)famous H-V mapping -- that possesses confined singularities despite being non-integrable. This fact cast serious doubt on results obtained by singularity confinement, even in the case where the use of the method was perfectly justified, for example in the deautonomisation procedure. The aim of the present work, already initiated in \cite{redeem}, is to show that singularity confinement can indeed, when implemented correctly, be a reliable and efficient discrete integrability detector. It goes without saying that in order to do this, one must explain the apparent paradox of non-integrable mappings with confined singularities. 

The origin of our present approach can be traced back to the algebro-geometric justification of the deautonomisation procedure we presented in \cite{prsa}. It was shown there that the constraints on the parameters obtained from singularity confinement were equivalent to the linear transformation induced, by the mapping, on part of the Picard group of the (family of) rational surfaces on which it can be regularised by blowing-up from $\Bbb{P}^1\times\Bbb{P}^1$. This phenomenon does not depend on the integrability of the mapping and we were thus led to the conjecture that for any confining mapping of the plane, the behaviour of the parameters of the mapping satisfying the confinement constraints, is governed by the linear action on the Picard group, obtained after successfully blowing-up the mapping. Moreover, as shown in \cite{take}, knowledge of the action on the Picard group allows one to calculate the algebraic entropy of the mapping rigorously. This observation was only one step away from the conjecture that if the deautonomisation of a confining mapping is sufficiently general so that its parameters depend on the largest eigenvalue of the linear map on the Picard group, the value of this eigenvalue and hence also the algebraic entropy of the mapping can be, so to speak, read-off directly from the confinement constraints themselves. The way to implement this prescription systematically is through the approach of full-deautonomisation we introduced in \cite{redeem}. The latter consists in considering a non-autonomous extension of a given mapping, not limited to the terms already present, but by considering all the terms that one can add to the mapping without modifying its singularity structure.

All known examples of confining non-integrable mappings, including the H-V mapping, have been explained within this new framework for singularity confinement. We were not only able to show why one does not expect these mappings to be integrable, but this was done by obtaining their algebraic entropy directly from the confinement constraints. This is a great simplification since the traditional way of computing the algebraic entropy of a confining mapping is either by painstakingly regularising the mapping through successive blow-ups, ultimately obtaining the associated action on the Picard group, or by iterating the mapping  up to sufficiently high orders so that one can establish a recursion relation for the degree growth with some degree of confidence. 

The notions of early and late confinement have been revisited in this paper. Early confinement, whenever possible, corresponds to the first possibility to confine and is already present in the autonomous case. This early confinement leads either to a trivial, usually linear, mapping or to a periodic one. Late confinement, on the other hand, can only exist in the non-autonomous case as it corresponds to a singularity pattern longer than that for the autonomous one. Since late confinement invariably leads to a non-integrable system, one might of course question its usefulness. However, as we have shown,  analysing a late confinement along with a normal one can help identify any inherent gauge freedom in the mapping, which, when treated in an improper way, might otherwise lead to wrong conclusions concerning its integrability. Of course, gauge freedom is totally transparent in an algebro-geometric treatment, but it is reassuring to be able to deal with this potential difficulty at the level of singularity analysis (after all, not everybody is a superhero who is able to find the proper gauge while standing on his head). Since a confining mapping, generically, has  an infinite number of periodically recurring late confinement opportunities and since at each step new confinement constraints will appear, there exist infinitely many possible values for the algebraic entropy for a deautonomised mapping, all of which can be simply obtained from the singularity confinement approach.

Finally, it is our belief that the singularity confinement criterion presented here, based on the full-deautonomisation method, can be cast upon a solid, rigorous, basis through an algebro-geometric approach and we intend to return to this question in the near future.

\section*{Acknowledgements}
RW would like to acknowledge support from the Japan Society for the Promotion of Science (JSPS), through the JSPS grant: KAKENHI grant number 15K04893. TM would also like to acknowledge support from JSPS through the Grant-in-Aid for Scientific Research  25-3088 and he wishes to thank the Graduate School of Mathematical Sciences of the University of Tokyo for support extended through its Program for Leading Graduate Schools, MEXT, Japan.

\section*{Appendix}
The eight base-points $Q_n^{(i)}$ ($i=1\cdots,8$) needed in the regularisation of the indeterminate point $(s_n, y_n) = (0, 0)$ for the mapping $\phi_n$ (as defined by \eqref{phimap}), together with the coordinate charts introduced in each blow-up and the resulting expressions for the mapping: 
\begin{itemize}
\item[i)] blow-up at $Q_n^{(1)}\!:(s_n, y_n) = (0, 0)$~;~coordinate chart :
\begin{gather}
(s_n, y_n) \leftarrow \left( s_n, \frac{y_n}{s_n} \right) \cup \left( \frac{s_n}{y_n}, y_n \right)
\quad\Rightarrow\quad y_{n+1} = \frac{1}{y^4_n \frac{s_n}{y_n}} \left( - y^3_n + b_n y^3_n \frac{s_n}{y_n} + a_n \frac{s_n}{y_n} \right)
\end{gather}
\item[ii)] blow-up at $Q_n^{(2)}\!:\left( \frac{s_n}{y_n}, y_n \right) = (0, 0)$~;~coordinate chart :
\begin{gather}
\left( \frac{s_n}{y_n}, y_n \right) \leftarrow \left( \frac{s_n}{y_n}, \frac{y^2_n}{s_n} \right) \cup \left( \frac{s_n}{y^2_n}, y_n \right)\quad
\Rightarrow\quad y_{n+1} = \frac{1}{y^4_n \frac{s_n}{y^2_n}} \left( - y^2_n + b_n y^3_n \frac{s_n}{y^2_n} + a_n \frac{s_n}{y^2_n} \right)
\end{gather}
\item[iii)] blow-up at $Q_n^{(3)}\!:\left( \frac{s_n}{y^2_n}, y_n \right) = (0, 0)$~;~coordinate chart :
\begin{gather}
\left( \frac{s_n}{y^2_n}, y_n \right) \leftarrow \left( \frac{s_n}{y^2_n}, \frac{y^3_n}{s_n} \right) \cup \left( \frac{s_n}{y^3_n}, y_n \right)\quad
\Rightarrow\quad y_{n+1} = \frac{1}{y^4_n \frac{s_n}{y^3_n}} \left( - y_n + b_n y^3_n \frac{s_n}{y^3_n} + a_n \frac{s_n}{y^3_n} \right)
\end{gather}
\item[iv)] blow-up at $Q_n^{(4)}\!:\left( \frac{s_n}{y^3_n}, y_n \right) = (0, 0)$~;~coordinate chart :
\begin{gather}
\left( \frac{s_n}{y^3_n}, y_n \right) \leftarrow \left( \frac{s_n}{y^3_n}, \frac{y^4_n}{s_n} \right) \cup \left( \frac{s_n}{y^4_n}, y_n \right)\quad
\Rightarrow\quad y_{n+1} = \frac{1}{y^4_n \frac{s_n}{y^4_n}} \left( - 1 + b_n y^3_n \frac{s_n}{y^4_n} + a_n \frac{s_n}{y^4_n} \right)
\end{gather}
\item[v)] blow-up at $Q_n^{(5)}\!:\left( \frac{s_n}{y^4_n}, y_n \right) = \left( \frac{1}{a_n}, 0 \right)$~;~coordinate chart :
\begin{gather}
\left( \frac{s_n}{y^4_n} - \frac{1}{a_n}, y_n \right) \leftarrow \left( \frac{s_n}{y^4_n} - \frac{1}{a_n}, \frac{y_n}{\frac{s_n}{y^4_n} - \frac{1}{a_n}} \right)\cup \left( \frac{1}{y_n} \left( \frac{s_n}{y^4_n} - \frac{1}{a_n} \right), y_n \right)\nonumber\\
\Rightarrow\quad y_{n+1} = \frac{1}{y^3_n \frac{s_n}{y^4_n}} \left( \frac{a_n}{y_n} \left( \frac{s_n}{y^4_n} - \frac{1}{a_n} \right) + b_n y^2_n \frac{s_n}{y^4_n} \right)
\end{gather}
\item[vi)] blow-up at $Q_n^{(6)}\!:\left( \frac{1}{y_n} \left( \frac{s_n}{y^4_n} - \frac{1}{a_n} \right), y_n \right) = (0, 0)$~;~coordinate chart :
\begin{gather}
\left( \frac{1}{y_n} \left( \frac{s_n}{y^4_n} - \frac{1}{a_n} \right), y_n \right) \leftarrow \left( \frac{1}{y_n} \left( \frac{s_n}{y^4_n} - \frac{1}{a_n} \right), \frac{y^2_n}{\frac{s_n}{y^4_n} - \frac{1}{a_n}} \right) \cup \left( \frac{1}{y^2_n} \left( \frac{s_n}{y^4_n} - \frac{1}{a_n} \right), y_n \right)\nonumber\\
\Rightarrow\quad y_{n+1} = \frac{1}{y^2_n \frac{s_n}{y^4_n}} \left( \frac{a_n}{y^2_n} \left( \frac{s_n}{y^4_n} - \frac{1}{a_n} \right) + b_n y_n \frac{s_n}{y^4_n} \right)
\end{gather}
\item[vii)] blow-up at $Q_n^{(7)}\!:\left( \frac{1}{y^2_n} \left( \frac{s_n}{y^4_n} - \frac{1}{a_n} \right), y_n \right) = (0, 0)$~;~coordinate chart :
\begin{gather}
\left( \frac{1}{y^2_n} \left( \frac{s_n}{y^4_n} - \frac{1}{a_n} \right), y_n \right) \leftarrow \left( \frac{1}{y^2_n} \left( \frac{s_n}{y^4_n} - \frac{1}{a_n} \right), \frac{y^3_n}{\frac{s_n}{y^4_n} - \frac{1}{a_n}} \right) \cup \left( \frac{1}{y^3_n} \left( \frac{s_n}{y^4_n} - \frac{1}{a_n} \right), y_n \right)\nonumber\\
\Rightarrow\quad y_{n+1} = \frac{1}{y_n \frac{s_n}{y^4_n}} \left( \frac{a_n}{y^3_n} \left( \frac{s_n}{y^4_n} - \frac{1}{a_n} \right) + b_n \frac{s_n}{y^4_n} \right)
\end{gather}
\item[viii)] blow-up at $Q_n^{(8)}\!:\left( \frac{1}{y^3_n} \left( \frac{s_n}{y^4_n} - \frac{1}{a_n} \right), y_n \right) = \left( - \frac{b_n}{a^2_n}, 0 \right)$~;~coordinate chart :
\begin{gather}
\left( \frac{1}{y^3_n} \left( \frac{s_n}{y^4_n} - \frac{1}{a_n} \right) + \frac{b_n}{a^2_n}, y_n \right) \leftarrow \left( \frac{1}{y^3_n} \left( \frac{s_n}{y^4_n} - \frac{1}{a_n} \right) + \frac{b_n}{a^2_n}, \frac{y_n}{\frac{1}{y^3_n} \left( \frac{s_n}{y^4_n} - \frac{1}{a_n} \right) + \frac{b_n}{a^2_n}} \right)\nonumber\\
\hskip7cm \cup \left( \frac{1}{y_n} \left( \frac{1}{y^3_n} \left( \frac{s_n}{y^4_n} - \frac{1}{a_n} \right) + \frac{b_n}{a^2_n} \right), y_n \right)\nonumber\\
\Rightarrow\quad y_{n+1} = \frac{1}{\frac{s_n}{y^4_n}} \left( \frac{a_n}{y_n} \left( \frac{1}{y^3_n} \left( \frac{s_n}{y^4_n} - \frac{1}{a_n} \right) + \frac{b_n}{a^2_n} \right) + b_n \left( \frac{s_n}{y^4_n} - \frac{1}{a_n} \right) \right).
\end{gather}
\end{itemize}

The base-points and coordinate charts needed in the successive blow-ups of the indeterminate point $(x_{n+1}, t_{n+1}) = (0, 0)$ of $\phi_n^{-1}$ :
\begin{itemize}
\item[i)] blow-up at $P_{n+1}^{(1)}\!:(x_{n+1}, t_{n+1}) = (0, 0)$~;~coordinate chart :
\begin{gather}
(x_{n+1}, t_{n+1}) \leftarrow \left( x_{n+1}, \frac{t_{n+1}}{x_{n+1}} \right) \cup \left( \frac{x_{n+1}}{t_{n+1}}, t_{n+1} \right)
\end{gather}
\item[ii)] blow-up at $P_{n+1}^{(2)}\!:\left( x_{n+1}, \frac{t_{n+1}}{x_{n+1}} \right) = (0, 0)$~;~coordinate chart :
\begin{gather}
\left( x_{n+1}, \frac{t_{n+1}}{x_{n+1}} \right) \leftarrow \left( x_{n+1}, \frac{t_{n+1}}{x^2_{n+1}} \right)\cup \left( \frac{x^2_{n+1}}{t_{n+1}}, \frac{t_{n+1}}{x_{n+1}} \right)
\end{gather}
\item[iii)] blow-up at $P_{n+1}^{(3)}\!:\left( x_{n+1}, \frac{t_{n+1}}{x^2_{n+1}} \right) = (0, 0)$~;~coordinate chart :
\begin{gather}
\left( x_{n+1}, \frac{t_{n+1}}{x^2_{n+1}} \right) \leftarrow \left( x_{n+1}, \frac{t_{n+1}}{x^3_{n+1}} \right)\cup \left( \frac{x^3_{n+1}}{t_{n+1}}, \frac{t_{n+1}}{x^2_{n+1}} \right)
\end{gather}
\item[iv)] blow-up at $P_{n+1}^{(4)}\!:\left( x_{n+1}, \frac{t_{n+1}}{x^3_{n+1}} \right) = (0, 0)$~;~coordinate chart :
\begin{gather}
\left( x_{n+1}, \frac{t_{n+1}}{x^3_{n+1}} \right) \leftarrow \left( x_{n+1}, \frac{t_{n+1}}{x^4_{n+1}} \right) \cup \left( \frac{x^4_{n+1}}{t_{n+1}}, \frac{t_{n+1}}{x^3_{n+1}} \right)
\end{gather}
\item[v)] blow-up at $P_{n+1}^{(5)}\!:\left( x_{n+1}, \frac{t_{n+1}}{x^4_{n+1}} \right) = \left( 0, \frac{1}{a_n} \right)$~;~coordinate chart :
\begin{gather}
\left( x_{n+1}, \frac{t_{n+1}}{x^4_{n+1}} - \frac{1}{a_n} \right) \leftarrow \left( x_{n+1}, \frac{1}{x_{n+1}} \left( \frac{t_{n+1}}{x^4_{n+1}} - \frac{1}{a_n} \right) \right) \cup \left( \frac{x_{n+1}}{\frac{t_{n+1}}{x^4_{n+1}} - \frac{1}{a_n}}, \frac{t_{n+1}}{x^4_{n+1}} - \frac{1}{a_n} \right)
\end{gather}
\item[vi)] blow-up at $P_{n+1}^{(6)}\!:\left( x_{n+1}, \frac{1}{x_{n+1}} \left( \frac{t_{n+1}}{x^4_{n+1}} - \frac{1}{a_n} \right) \right) = (0, 0)$~;~coordinate chart :
\begin{gather}
\left( x_{n+1}, \frac{1}{x_{n+1}} \left( \frac{t_{n+1}}{x^4_{n+1}} - \frac{1}{a_n} \right) \right) \leftarrow \left( x_{n+1}, \frac{1}{x^2_{n+1}} \left( \frac{t_{n+1}}{x^4_{n+1}} - \frac{1}{a_n} \right) \right)\nonumber\\
\hskip8cm    \cup \left( \frac{x^2_{n+1}}{\frac{t_{n+1}}{x^4_{n+1}} - \frac{1}{a_n}}, \frac{1}{x_{n+1}} \left( \frac{t_{n+1}}{x^4_{n+1}} - \frac{1}{a_n} \right) \right)
\end{gather}
\item[vii)] blow-up at $P_{n+1}^{(7)}\!:\left( x_{n+1}, \frac{1}{x^2_{n+1}} \left( \frac{t_{n+1}}{x^4_{n+1}} - \frac{1}{a_n} \right) \right) = (0, 0)$~;~coordinate chart :
\begin{gather}
\left( x_{n+1}, \frac{1}{x^2_{n+1}} \left( \frac{t_{n+1}}{x^4_{n+1}} - \frac{1}{a_n} \right) \right) \leftarrow \left( x_{n+1}, \frac{1}{x^3_{n+1}} \left( \frac{t_{n+1}}{x^4_{n+1}} - \frac{1}{a_n} \right) \right)\nonumber\\
\hskip8cm    \cup \left( \frac{x^3_{n+1}}{\frac{t_{n+1}}{x^4_{n+1}} - \frac{1}{a_n}}, \frac{1}{x^2_{n+1}} \left( \frac{t_{n+1}}{x^4_{n+1}} - \frac{1}{a_n} \right) \right)
\end{gather}
\item[viii)] blow-up at $P_{n+1}^{(8)}\!:\left( x_{n+1}, \frac{1}{x^3_{n+1}} \left( \frac{t_{n+1}}{x^4_{n+1}} - \frac{1}{a_n} \right) \right) = \left( 0, - \frac{b_n}{a^2_n} \right)$~;~coordinate chart :
\begin{gather}
\left( x_{n+1}, \frac{1}{x^3_{n+1}} \left( \frac{t_{n+1}}{x^4_{n+1}} - \frac{1}{a_n} \right) + \frac{b_n}{a^2_n} \right) \leftarrow \left( x_{n+1}, \frac{1}{x_{n+1}} \left( \frac{1}{x^3_{n+1}} \left( \frac{t_{n+1}}{x^4_{n+1}} - \frac{1}{a_n} \right) + \frac{b_n}{a^2_n} \right) \right)\nonumber\\
\hskip5cm    \cup \left( \frac{x_{n+1}}{\frac{1}{x^3_{n+1}} \left( \frac{t_{n+1}}{x^4_{n+1}} - \frac{1}{a_n} \right) + \frac{b_n}{a^2_n}}, \frac{1}{x^3_{n+1}} \left( \frac{t_{n+1}}{x^4_{n+1}} - \frac{1}{a_n} \right) + \frac{b_n}{a^2_n} \right)
\end{gather}
\end{itemize}


\end{document}